\documentclass[epj]{svjour}
\usepackage{graphicx}
\usepackage{amsmath,amssymb,bm}
\usepackage{color,soul}
\usepackage{mathabx}
\begin{document}
\title{Conventional and inverse magnetocaloric and electrocaloric effects of a  mixed spin-(1/2, 1) Heisenberg dimer }
\author{Hana Vargov\'a\inst{1,}\thanks{\emph{Corresponding author:} hcencar@saske.sk}
\and Jozef Stre\v{c}ka\inst{2}
%
}                     
%
%
\institute{Institute of Experimental Physics, Slovak Academy of Sciences, Watsonova 47, 040 01 Ko\v {s}ice, Slovakia \and Department of Theoretical Physics and Astrophysics, Faculty of Science, P. J. \v{S}af\'{a}rik University,\\ Park 
Angelinum 9, 040 01 Ko\v{s}ice, Slovakia}
\date{Received: date / Revised version: date}
%
\abstract{
The mixed spin-(1/2, 1) Heisenberg dimer accounting for two different Land\'e $g$-factors is exactly examined in presence of external magnetic and electric field by considering  exchange as well as   uniaxial single-ion anisotropies.  Rigorously calculated ground-state phase diagrams affirm existence of three different types of zero-temperature phase transitions accompanied with a non-zero value of a residual entropy.  Presence of a magnetoelectric effect accounted within Katsura-Nagaosa-Balatsky mechanism is demonstrated through the analyzis of the magnetization and  dielectric polarization in response to both external fields.  The analyzis of two basic magnetocaloric characteristics, the adiabatic change of temperature and  the isothermal entropy change, achieved upon  variation of external fields, are exactly calculated in order to investigate the (multi)caloric behavior. The obtained results confirm existence of both conventional as well as inverse magnetocaloric effects.  Utilizing the refrigeration capacity coefficient it is found that the application of an electric field  during the adiabatic demagnetization process may lead to an enhancement of cooling performance  in the region of conventional magnetocaloric effect. On the other hand, a sufficiently large electric field can reduce an inverse caloric effect provided that  the electric-field-induced transition from the fully to partially polarized state is realized.
\PACS{
      {PACS-key}{discribing text of that key}   \and
      {PACS-key}{discribing text of that key}
     } 
} 
\maketitle
\section{Introduction}
\label{intro}
Low-dimensional  quantum spin models traditionally belong to the most intensively studied magnetic systems. A low-dimensionality promoting extraordinary strong quantum fluctuations is at an origin of    considerable diversity of  unconventional properties such as a quantum entanglement~\cite{Wootters,Souza,Horodecki,Rojas2012,Cenci20,Cenci21,Varga21}, existence of plateaus or quasi-plateaus in  magnetization curves~\cite{Honecker,Ohanyan2015,Cenci18}, quantum spin-liquid state~\cite{Lacroix,Imambekov} or  magnetoelectric~\cite{Sharples,Baran,Cenci98,Ohanyan20} effect, which makes low-dimensional quantum spin systems  very promising for various technological applications in modern smart devices. Beside to this, an enormous proliferation in an investigation of  low-dimensional quantum materials may be connected to their utilization in quantum communication and quantum information processing~\cite{Markham,Matthews,Rahaman}. 

Caloric effects, which are characterized as temperature changes of a physical system upon  variation of applied external magnetic  and/or electric field, are further extraordinary features of low-dimensional quantum spin systems referred to as magnetocaloric and electrocaloric phenomenon~\cite{Fu,Oliveira,Torrico,Szalowski2019,Mischenko,Moya14,Szalowski18}. An effort aimed at deeper understanding of both unconventional caloric processes has stimulated vigorous researcher's activities,  bearing in mind the fact that the solid-state cooling/heating technologies are environmentally more friendly alternative with respect to a conventional vapor-cycle  refrigeration~\cite{Glanz}. In addition, the wide utilization of aforementioned phenomena in the ultra-low cryogenics, room temperature cooling or cooling of (micro)electronic components, makes  materials with a substantial caloric response very demanding for various spheres of a life. Despite the energetically and thus economically higher convenience of an electrocaloric effect (ECE), the magnetocaloric effect (MCE) was to date a more thoroughly  investigated~\cite{Ismail}. A few recent studies devoted to the concept of multicaloric materials argued that  the MCE could be maximized by an external electric field~\cite{Ursic,Qiao,Szalowski21}. Due to a complexity of processes responsible for existence of such intricate behavior there exists only a few theoretical studies dealing with the multicaloric behavior of low-dimensional quantum spin systems~\cite{Baran,Szalowski18,Zad,Ding,Richter,Fodouop}. For this reason, we would like to contribute to this novel research area with a goal to extend  knowledge about the maximization of caloric processes in solid-state magnetic insulators. 

It should be emphasized that  various caloric processes can be in general accompanied by  relieving or  consuming of an additional heat, which is macroscopically detectable as a heating or a cooling of a system environment. The conventional  caloric  effect refers to the positive isothermal entropy change and negative adiabatic temperature change achieved  upon the variation of an external field, while the negative isothermal entropy change and positive adiabatic temperature change is termed as an inverse caloric effect.  The inverse MCE was for instance reported for  different kinds of antiferromagnetic, ferrimagnetic or  paramagnetic  compounds~\cite{Ranke,Krenke1,Krenke2,Sandeman,Nayak,Biswas}, while   the inverse ECE has been observed in ferroelectric~\cite{Udin,Geng} and antiferroelectric materials specifically restricted to a certain temperature and field regions~\cite{Perantie,LeGoupil}. 
A coexistence of both conventional and inverse caloric phenomena is rarely also possible for advanced multifunctional materials with an extraordinary  magnetic phase diagram~\cite{Szalowski21,Derzhko,Pereira,Verkholyak,Szalowski,Galisova,Krishnamoorthi,Diop,Odaira,Burzo,Zhang2010,Reis}.
 In connection to this, the partial aim of this study is to examine the possibility to generate  different types of thermal changes (the cooling or heating) utilizing peculiar properties of heterogeneous low-dimensional quantum spin systems. 

From the theoretical point of view, the quantum Heisenberg model and  its diverse variants represent a good theoretical tool to study     low-dimensional quantum spin systems, which allow due to their relative simplicity  a rigorous examination of  quantum phenomena in their purest nature~\cite{Gatteschi,Kahn,Hieida,Dmitriev04,Hikihara,Hagemans,Caux,Menchyshyn,Brockmann,Chen,Bacq,Cu,Ivanov,Rojas2017,Rojas2019,Strecka20}.   Especially, the following works focusing on  the study of thermodynamic and magnetic properties of various Heisenberg dimers~\cite{Furrer,Haraldsen,Efremov,Whangbo} are rewarding to better understand   peculiar properties of single molecular magnets.  In the present paper we will focus our attention on one of the simplest  Heisenberg models, namely, a mixed spin-(1/2, 1) Heisenberg dimer, which involves two different types of magnetic ions.  The choice of the magnetically asymmetric system has been performed for two reasons. The first one originates from the fact, that  the mixed spin-(1/2, 1) Heisenberg dimer  very well approximates a distinct group of bimetallic low-dimensional molecular magnets like that reported in Refs.~\cite{Kahn,Gleizes,Hagiwara2,Koningsbruggen,Hagiwara,Pei}.  And the second one originates from the knowledge, that the  imbalance between magnetic ions leads to the modification of excitation spectra~\cite{Houchins},  which can be a conceptus of  different caloric properties. The application of the additional electric field on such simple quantum system in connection to the possible enhancement of MCE is a supplemental task of our theoretical analyzis.

The paper is organized as follows. In Sec.~\ref{model} we briefly introduce the model under the investigation along with  relevant quantities required for a study of ground-state and thermodynamic properties. The most interesting results  demonstrating the influence of both magnetic as well as electric field on the  ground-state spin arrangement, total magnetization, dielectric polarization and entropy are discussed in Sec.~\ref{results}. The dominant part of Sec.~\ref{results} is focused on the study of magnetocaloric properties  and their modulation caused by  the varied electric field. A summary of the most important findings are presented in Sec.~\ref{conclusion}.

\section{Model and Method}
\label{model}
Let us consider a mixed spin-(1/2, 1) Heisenberg dimer defined through the Hamiltonian 
\allowdisplaybreaks
\begin{align}
\hat{\cal H}&=J\left[\Delta(\hat{S}^x_1\hat{\mu}^x_2\!+\!\hat{S}^y_1\hat{\mu}^y_2)\!+\!\hat{S}^z_1\hat{\mu}^z_2\right]\!+\!D(\hat{\mu}^z_2)^2
\!-\!\mu_BB(g_1\hat{S}^z_1\!+\!g_2\hat{\mu}^z_2)\!+\!E(\hat{S}^x_1\hat{\mu}^y_2\!-\!\hat{S}^y_1\hat{\mu}^x_2).
\label{eq1}
\end{align} 
\begin{figure}[t!]
{\includegraphics[width=.25\textwidth,trim=0.5cm 6.8cm 16.8cm 19.3cm, clip]{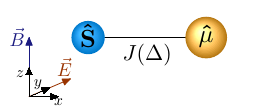}}
\caption{A schematic representation of a mixed spin-(1/2, 1) Heisenberg dimer defined through  the Hamiltonian~\eqref{eq1} and situated along  $x$-axis. Blue (small) ball illustrates the spin-1/2 magnetic ion and yellow (large) ball illustrates the spin-1 magnetic ion coupled through the exchange constant $J(\Delta)$. External magnetic field  ($\widearrow{B}$) is  applied along the $z$-axis, while the external electric field ($\widearrow{E}$) is applied along the  $y$-axis. }
\label{fig1}
\end{figure}
The symbols $\hat{S}^{\alpha}_1$ and $\hat{\mu}^{\alpha}_2$ ($\alpha\!=\!x,y,z$) denote  spatial components of spin-1/2 and 1 operators, the parameter $J$ characterizes the exchange interaction  with the  XXZ exchange anisotropy parameter $\Delta$ and the parameter  $D$ corresponds to  the uniaxial single-ion anisotropy   acting exclusively on the spin-1 magnetic ion.   The external magnetic field $\widearrow{B}\!=\!(0,0,B)$  applied along $z$-axis is responsible for existence of the standard Zeeman's term in the Hamiltonian (1), while  the external electric field applied along $y$-direction $\widearrow{E}\!=\!(0,E_y,0)$ is being responsible for other non-zero contribution denoted by the electric field energy $E$.  Both  external fields are perpendicular to a link connecting both spin species lying along $x$-axis (see Fig.~\ref{fig1}). Finally, $g_1$ and $g_2$ are Land\'e $g$-factors of the spin-1/2 and spin-1 magnetic ions, respectively, the symbol $\mu_B$ stands for Bohr magneton. The specific form of the  electric field energy follows from the inverse Dzyaloshinskii-Moriya mechanism referred to as Katsura-Nagaosa-Balatsky (KNB) mechanism~\cite{Katsura,Jia}, according to which the dielectric polarization  of the pair of interacting spins $\hat{\mathbf{S}}_1$ and $\hat{\bm{\mu}}_2$ is connected to  the following expression 
\begin{align}
\hat{\mathbf{P}}_{12}\!\varpropto\!{\widearrow{\rm e}}_{12}\!\times\!\hat{\mathbf{S}}_1\!\times\!\hat{\bm{\mu}}_2,
\label{eq2x}
\end{align}
where ${\widearrow{\rm e}}_{12}$ is the unit vector pointing from site 1 to site 2. In the present case  the magnetic ions are positioned along the $x$-axis, \textit{i.e.} $\widearrow{\rm e}_{12}\!=\!(1,0,0)$, and hence, the spatial components of the dielectric polarization according to the KNB mechanism  are quite simple
\begin{align}
\hat{P}_x\!=\!0,\hspace*{1cm} \hat{P}_y\!\varpropto\!\hat{S}^y_1\hat{\mu}^x_{2}\!-\!\hat{S}^x_1\hat{\mu}^y_{2},\hspace*{1cm}\hat{P}_z\!\varpropto\!\hat{S}^z_1\hat{\mu}^x_{2}\!-\!\hat{S}^x_1\hat{\mu}^z_{2}.
\label{eq3x}
\end{align}
It directly follows from Eq.~\eqref{eq3x} that a specific spatial orientation of  the external electric field applied along the $y$-axis leads to the particular form of the Hamiltonian~\eqref{eq1}, which exemplifies the energy contribution $E$ of the external electric field  on the mixed spin-(1/2, 1) Heisenberg dimer  by means of the effective Dzyaloshinskii-Moriya term $E(\hat{S}^x_1\hat{\mu}^y_2\!-\!\hat{S}^y_1\hat{\mu}^x_2)$.  In addition, the effective Dzyaloshinskii-Moriya term determines, up to the unimportant constant $\mu$, also the dielectric polarization $P\!\equiv\!\langle \hat{P}_y\rangle\!=\!\mu\langle \hat{S}^y_1\hat{\mu}^x_{2}\!-\!\hat{S}^x_1\hat{\mu}^y_{2}\rangle$, whereby the microscopic constant $\mu$ depends on quantum chemical features of the bond connecting two magnetic ions~\cite{Katsura,Jia}.

The matrix representation of the Hamiltonian (\ref{eq1}) constructed in a standard basis of eigenstates of $z$-components of both spin operators $\vert\varphi_i\rangle\!\equiv\!\vert S^z_1,\mu^z_2\rangle\!\in\!\{\vert\frac{1}{2},1\rangle,\vert\frac{1}{2},0\rangle,\vert\frac{1}{2},-1\rangle,\vert\!-\!\frac{1}{2},1\rangle,\vert\!-\!\frac{1}{2},0\rangle,\vert\!-\!\frac{1}{2},-1\rangle\}$ takes a  relative simple (sparse) form 
\begin{align}
&\left\langle \varphi_j \right\vert \hat{\cal H} \left\vert\varphi_i \right\rangle
=\left(
\begin{array}{cccccc}
{\cal H}_{11}& 0 & 0& 0 & 0 & 0\\
0 &{\cal H}_{22} & 0 &{\cal H}_{24}& 0 & 0 \\
0 & 0 & {\cal H}_{33} & 0 & {\cal H}_{35} & 0\\
0 & {\cal H}_{42} & 0 & {\cal H}_{44} & 0 & 0\\
0 & 0 & {\cal H}_{53}& 0 & {\cal H}_{55} & 0  \\
0 & 0& 0 & 0 & 0 & {\cal H}_{66}
\end{array}\right)
\label{eq4x}
\end{align}
with the following six  diagonal elements ${\cal H}_{ii}$ ($i\!=\!1,2,\dots,6$) 
 \allowdisplaybreaks
\begin{align}
{\cal H}_{11}\!&=\!\frac{1}{2}\left[J\!+\!2D\!-\!(h_1\!+\!2h_2)\right],
\hspace*{1cm}
{\cal H}_{22}\!=\!-{\cal H}_{55}\!=\!-\frac{h_1}{2},
\hspace*{1cm}
{\cal H}_{33}\!=\!-\frac{1}{2}\left[J\!-\!2D\!+\!(h_1\!-\!2h_2)\right],
\nonumber\\
{\cal H}_{44}\!&=\!-\frac{1}{2}\left[J\!-\!2D\!-\!(h_1\!-\!2h_2)\right],
\hspace*{0.75cm}
{\cal H}_{66}\!=\!\frac{1}{2}\left[J\!+\!2D\!+\!(h_1\!+\!2h_2)\right].
\label{eq5x}
\end{align}
In  above expressions  two new parameters $h_{1,2}\!=\!g_{1,2}\mu_BB$ have been introduced with the aim to simplify the mathematical notation of further  relevant quantities.
 Four remaining non-zero off-diagonal  elements  are complex   due to  the last term of Eq.~(\ref{eq1}), which has character of antisymmetric Dzyaloshinskii-Moriya interaction originating from the external electric field applied along the $y$-axis. Thus, the non-zero off-diagonal terms can be unambiguously characterized by the argument $\varphi\!=\!\arctan \left(\frac{E}{J\Delta}\right)$ through the following formulas
\begin{align}
{\cal H}_{24}\!&=\!{\cal H}_{35}\!=\!\frac{{\rm e}^{i\varphi}}{\sqrt{2}}\sqrt{(J\Delta)^2\!+\!E^2},
\hspace*{1cm}
{\cal H}_{42}\!=\!{\cal H}_{53}\!=\!\frac{{\rm e}^{-i\varphi}}{\sqrt{2}}\sqrt{(J\Delta)^2\!+\!E^2}.
\label{eq6x}
\end{align}
 After a straightforward diagonalization of the Hamiltonian (\ref{eq4x})  one can obtain a complete energy spectrum involving six different eigenvalues and respective eigenvectors
\begin{align}
\varepsilon_{1,2}=&\frac{1}{2}\left[J\!+\!2D\!\mp\!(h_1\!+\!2h_2)\right],
\hspace*{6.5cm}
\vert\psi_{1,2}\rangle=\vert\!\pm\!\tfrac{1}{2},\pm1\rangle, 
\label{eq7x}\\\nonumber\\
\varepsilon_{3,4}=&-\frac{1}{4}\left(J\!-\!2D\!+\!2h_2\right)
\!\mp\!\frac{1}{4}\sqrt{\left[J\!-\!2D\!-\!2(h_1\!-\!h_2)\right]^2\!+\!8\left[(J\Delta)^2\!+\!E^2\right]},
\hspace*{0.5cm}
\vert\psi_{3,4}\rangle={\rm e}^{i\varphi}|c_1^{\mp}|\vert\tfrac{1}{2},0\rangle\!\mp\!|c_1^{\pm}|\vert\!-\!\tfrac{1}{2},1\rangle,
\label{eq8x}\\\nonumber\\
\varepsilon_{5,6}=&-\frac{1}{4}\left(J\!-\!2D\!-\!2h_2\right)
\!\mp\!\frac{1}{4}\sqrt{\left[J\!-\!2D\!+\!2(h_1\!-\!h_2)\right]^2\!+\!8\left[(J\Delta)^2\!+\!E^2\right]},
\hspace*{0.5cm}
\vert\psi_{5,6}\rangle={\rm e}^{i\varphi}|c_2^{\pm}|\vert\tfrac{1}{2},-1\rangle\!\mp\!|c_2^{\mp}|\vert\!-\!\tfrac{1}{2},0\rangle.
\label{eq9x}
\end{align} 
The modulus of complex coefficients of four nonseparable quantum entangled states  $\vert\psi_{3,4}\rangle$ and $\vert\psi_{5,6}\rangle$  are explicitly defined by the following relations
\begin{align}
|c_1^{\pm}|&=\frac{1}{\sqrt{2}}\sqrt{1\!\pm\!\frac{J\!-\!2D\!-\!2(h_1\!-\!h_2)}{\sqrt{\left[J\!-\!2D\!-\!2(h_1\!-\!h_2)\right]^2\!+\!8\left[(J\Delta)^2\!+\!E^2\right]}}
}, \hspace*{0.5cm}
|c_2^{\pm}|=\frac{1}{\sqrt{2}}\sqrt{1\!\pm\!\frac{J\!-\!2D\!+\!2(h_1\!-\!h_2)}{\sqrt{\left[J\!-\!2D\!+\!2(h_1\!-\!h_2)\right]^2\!+\!8\left[(J\Delta)^2\!+\!E^2\right]}}
}.
\label{eq10x}
\end{align} 

With the help of a complete set of energy eigenvalues~\eqref{eq7x}-\eqref{eq9x}  one may  obtain an exact analytical expression for the partition function ${\cal Z}$
\begin{align}
{\cal Z}\!=\!\sum_{i=1}^6 {\rm e}^{- \varepsilon_i/k_BT}
\label{eq11x}
\end{align} 
where $k_B$ is a Boltzmann's constant and $T$ is the absolute temperature. The exact analytic expression for the partition function is for brevity explicitly given  in Appendix, Eq.~\eqref{eq11xa}. 
Having an explicit expression for the partition function (\ref{eq11x}) one can easily derive the free energy $F\!=\!-k_BT\ln {\cal Z}$ and subsequently all thermodynamic functions relevant for the analyzis of  magnetoelectric properties. In particular, the total magnetization $m$ normalized with respect to the saturation magnetization $m_s$ 
\begin{align}
\frac{m}{m_s}&\!=\!-\frac{\partial F}{\partial \mu_BB}
\label{eq12x}
\end{align}
and the dielectric polarization $P$  follows from the formula
\begin{align}
P\!&=\!-\mu\frac{\partial F}{\partial E},
\label{eq13x}
\end{align}
  which will be from here onward considered  in dimensionless units by setting the microscopic constant $\mu\!=\!1$. Analyzing the explicit form of both formulas, explicitly listed in Appendix as Eq.~\eqref{eq12xa} and Eq.~\eqref{eq13xa}, one immediately deduces  that the magnetization and dielectric polarization become zero in absence of external magnetic and  electric fields, respectively, what immediately precludes existence of a spontaneous multiferroic behavior.
 From the point of view of ECE and MCE properties the most essential quantity is the magnetic entropy $S$, defined as
\begin{align}
\displaystyle
S&=-\frac{\partial F}{\partial T},
\label{eq14x}
\end{align}
 whose exact analytic formula is for completeness given in  Appendix, Eq.~\eqref{eq14xa}.
\section{Results and discussion}
\label{results}
In this section, we   will proceed to a discussion of the most interesting results obtained for   the mixed spin-(1/2, 1) Heisenberg dimer in a presence of the external magnetic and electric fields with the particular emphasis laid on conventional and inverse caloric effects.  Our analyzis will be restricted to the particular case with the antiferromagnetic fully isotropic exchange coupling $J\!>\!0$, $\Delta\!=\!1$, which however reflects all generic features of the investigated quantum spin model even for a more general case with $\Delta\!\neq\!1$ as well. For simplicity, the  Land\'e $g$-factor of the spin-1/2 magnetic ion is fixed to the value $g_1\!=\!2$, whereas the Land\'e $g$-factor of the spin-1 magnetic ion $g_2$ will be varied.

\subsection{Ground state}

\begin{figure*}[h!]
{\includegraphics[width=.45\textwidth,trim=0.5cm 7.5cm 1.8cm 8.cm, clip]{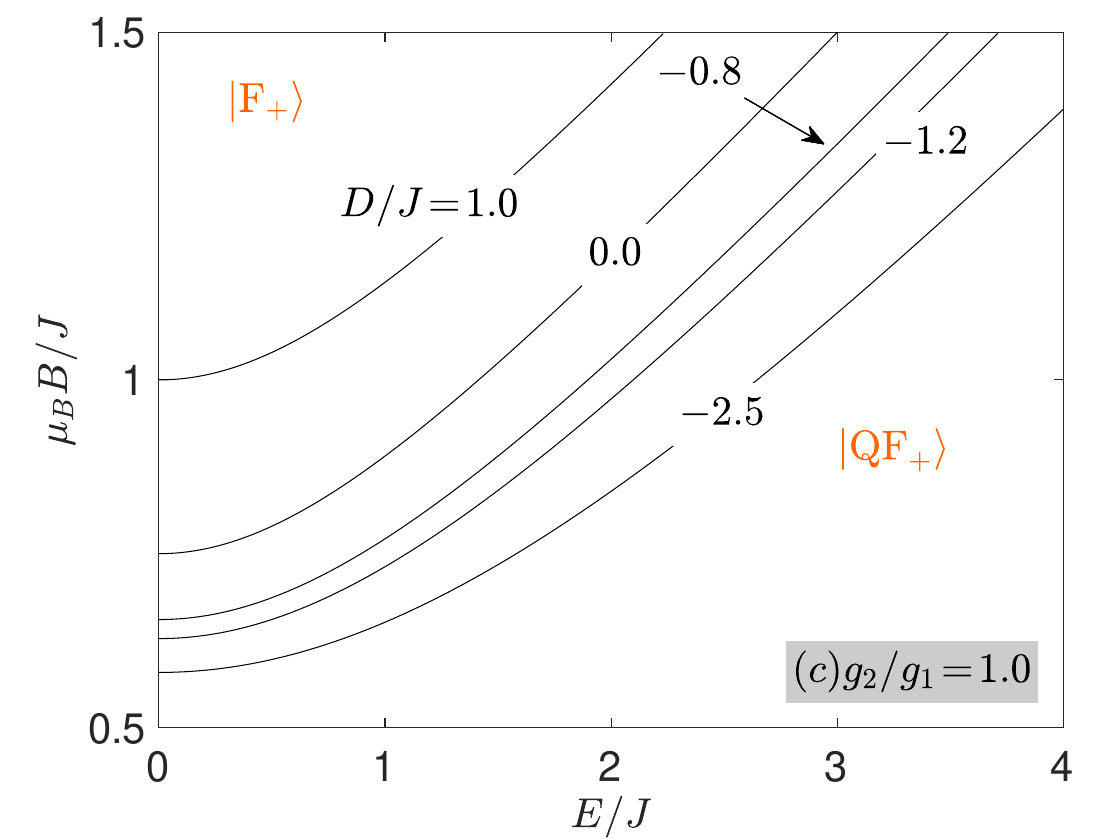}}
{\includegraphics[width=.45\textwidth,trim=0.5cm 7.5cm 1.8cm 8.cm, clip]{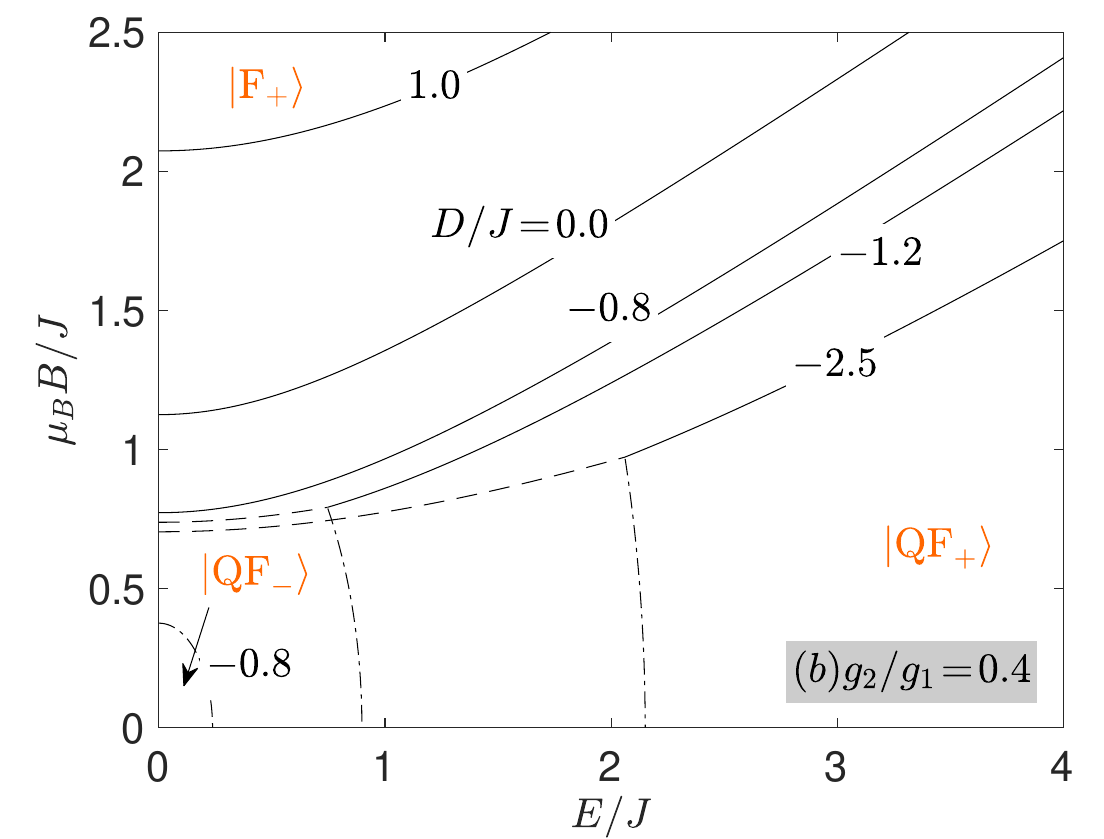}}
\caption{Ground-state phase diagrams in the $E/J\!-\!\mu_BB/J$ plane  for  a few selected values of the uniaxial single-ion anisotropy $D/J$,  $\Delta\!=\!1.0$ and the two different ratios of the $g$-factors:  $(a)$ $g_2/g_1\!=\!1$ and $(b)$ $g_2/g_1\!=\!0.4$. The displayed lines in all panels correspond to the phase boundaries: $\vert {\rm QF}_{+}\rangle$-$\vert {\rm F}_+\rangle$ (solid lines),  $\vert {\rm QF}_{-}\rangle$-$\vert {\rm F}_+\rangle$ (dashed lines) and $\vert {\rm QF}_{-}\rangle$-$\vert {\rm QF}_{+}\rangle$  (dashed-dotted lines), respectively. }
\label{fig2}
\end{figure*}

It turns out that the mixed spin-(1/2, 1) Heisenberg dimer has in total three different   ground states, which can be   characterized via the following eigenvectors and eigenenergies 
\begin{align}
|{\rm F}_{\pm}\rangle&=
|\!\pm\!\tfrac{1}{2},\pm1\rangle,
\hspace*{3.5cm}
\varepsilon_{\rm F_{\pm}}=\frac{1}{2}\left[J\!+\!2D\!\mp\!(h_1\!+\!2h_2)\right],
\nonumber\\
|{\rm QF_{+}}\rangle&={\rm e}^{i\varphi}|c_1^{-}|\vert\tfrac{1}{2},0\rangle\!-\!|c_1^{+}|\vert\!-\!\tfrac{1}{2},1\rangle,
\hspace*{0.6cm}
\varepsilon_{\rm QF_{+}}=-\frac{1}{4}\left(J\!-\!2D\!+\!2h_2\right)
\!-\!\frac{1}{4}\sqrt{\left[J\!-\!2D\!-\!2(h_1\!-\!h_2)\right]^2\!+\!8\left[(J\Delta)^2\!+\!E^2\right]},
\nonumber\\
|{\rm QF_{-}}\rangle&=|c_2^{-}|\vert\!-\!\tfrac{1}{2},0\rangle\!-\! {\rm e}^{i\varphi}|c_2^{+}|\vert\tfrac{1}{2},-1\rangle,
\hspace*{0.3cm}
\varepsilon_{\rm QF_{-}}=-\frac{1}{4}\left(J\!-\!2D\!-\!2h_2\right)
\!-\!\frac{1}{4}\sqrt{\left[J\!-\!2D\!+\!2(h_1\!-\!h_2)\right]^2\!+\!8\left[(J\Delta)^2\!+\!E^2\right]}.
\nonumber
\end{align} 
Evidently, the  fully polarized  ferromagnetic ground state  $\vert {\rm F}_{\pm}\rangle$ is two-fold degenerate in absence of an external magnetic field, but  an arbitrary small but non-zero magnetic field lift this two-fold degeneracy as both  spins are fully polarized to the magnetic-field direction. The quantum ferrimagnetic ground states $\vert {\rm QF}_{\pm}\rangle$ differing from each other by the sign of a $z$-component of the total spin (the '$+$' sign for the $\vert {\rm QF}_{+}\rangle$ phase and the '$-$' sign for the $\vert {\rm QF}_{-}\rangle$ phase) are also energetically equivalent in the zero magnetic field. 
It should be emphasized that the obtained results are in a good correspondence with previous observations~\cite{Cenci20}, because  the  quantum ferrimagnetic ground state $\vert{\rm QF}_{-}\rangle$  emerges in the ground-state phase diagram just in the case of low enough $g_2/g_1$ ratio obeying the condition
\begin{align}
\frac{g_2}{g_1}\!&\leq\!\left\{1+\sqrt{1\!+\!8\frac{(J\Delta)^2\!+\!E^2}{(J\!-\!2D)^2}}\right\}^{-1}\!\lesssim\!\frac{1}{2},
\label{eq15x}
\end{align} 
which ensures a positive total magnetization in spite of the negative total spin.  In addition, the existence of the quantum ferrimagnetic phase $\vert{\rm QF}_{-}\rangle$ with the negative sign of the total spin is strongly determined by the specific value of an uniaxial single-ion anisotropy $D/J$ given   at least  by one of the following conditions
\begin{align}
&\varepsilon_{\rm QF_{-}}\!\leq\!\varepsilon_{\rm F_+}:D\!\leq\!\frac{1}{2}\left[2\left(h_1\!+\!h_2\right)\!-\!J\!-\!\frac{(J\Delta)^2\!+\!E^2}{(2h_2\!-\!J)} \right],
\hspace*{1.cm}
\varepsilon_{\rm QF_{-}}\!\leq\!\varepsilon_{\rm QF_{+}}:|J\!-\!2D|\!\geq\!
2h_2\sqrt{1\!+\!\frac{2[(J\Delta)^2\!+\!E^2]}{h_1(h_1\!-\!2h_2)}}.
\label{eq16x}
\end{align} 
Fig.~\ref{fig2} illustrates the typical ground-state phase diagrams in the $E/J-\mu_BB/J$ plane for the ratio $g_2/g_1$ above (Fig.~\ref{fig2}$(a)$) and below (Fig.~\ref{fig2}$(b)$) the critical ratio of the Land\'e $g$-factors, $g_2/g_1\!\approx\!1/2$. 
 The ground-state phase diagram has for each $g_2/g_1\!>\!1/2$ a very simple structure  with just two phases, $\vert{\rm QF}_+\rangle$ and $\vert{\rm F}_+\rangle$,  which have the equivalent energies at the magnetic fields restricted by the condition 
\begin{align}
\vert {\rm QF}_{+}\rangle-\vert {\rm F}_+\rangle: \mu_BB\!=\!\frac{1}{4}\left[ \left(\frac{J\!+\!2D}{g_2}\!+\!\frac{2J}{g_1}\right)\!+\!\sqrt{ \left(\frac{J\!+\!2D}{g_2}\!-\!\frac{2J}{g_1}\right)^2\!+\!\frac{8\left[(J\Delta)^2\!+\!E^2\right]}{g_1g_2}}\right].
\label{eq17x}
\end{align}
It is obvious from Eq.~\eqref{eq17x}   that the increasing $g_2/g_1$ ratio and/or  decreasing value of the uniaxial single-ion anisotropy $D/J$ support the effect of the magnetic field to reorient  magnetic moments of both ions into its direction. The typical structure of the ground-state phase diagram for $g_2/g_1<1/2$ is presented  in Fig.~\ref{fig2}$(b)$. Above the threshold magnetic field  $\mu_BB_t/J$ the ground-state phase diagram is quite reminiscent of the one discussed above, whereas below this threshold magnetic field both  quantum ferrimagnetic phases  $\vert{\rm QF}_{\pm}\rangle$ coexist together whenever the  condition is met
\begin{align}
\vert {\rm QF}_{-}\rangle-\vert {\rm QF}_+\rangle: \mu_BB\!=\!\frac{1}{2}\sqrt{\left(\frac{J\!-\!2D}{g_2}\right)^2\!-\!\frac{8[\left(J\Delta\right)^2\!+\!E^2]}{g_1(g_1-2g_2)}}.
\label{eq18x}
\end{align}
Obviously, the decreasing uniaxial single-ion anisotropy enlarges the stability region of the peculiar quantum ferrimagnetic phase with the negative total spin $\vert{\rm QF}_-\rangle$  separated from the fully polarized $\vert{\rm QF}_+\rangle$ phase by the condition

\begin{align}
\vert {\rm QF}_{-}\rangle-\vert {\rm F}_+\rangle: \mu_BB\!=\! \frac{1}{4}\left[ \left(\frac{J\!+\!2D}{(g_1\!+\!g_2)}\!+\!\frac{J}{g_2}\right)\right.
\!+\!\left.\sqrt{ \left(\frac{J\!+\!2D}{(g_1\!+\!g_2)}\!-\!\frac{J}{g_2}\right)^2\!+\!\frac{4\left[\left(J\Delta\right)^2\!+\!E^2\right]}{g_2(g_1\!+\!g_2)}}\right].
\label{eq19x}
\end{align}

\subsection{Magnetoelectric effect}
\begin{figure*}[t!]
{\includegraphics[width=0.33\textwidth,trim=0.9cm 7cm 0.5cm 7.5cm,, clip]{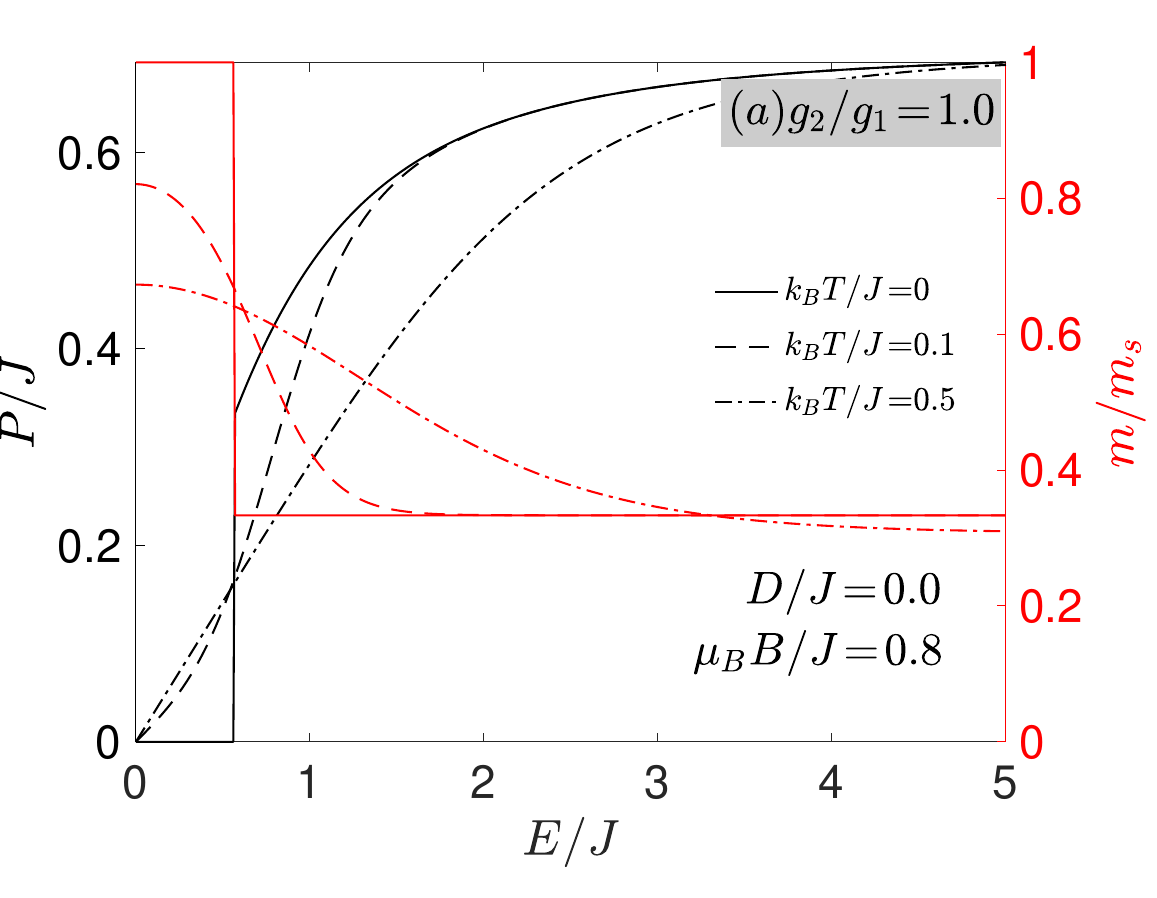}}
{\includegraphics[width=0.33\textwidth,trim=0.9cm 7cm 0.5cm 7.5cm,, clip]{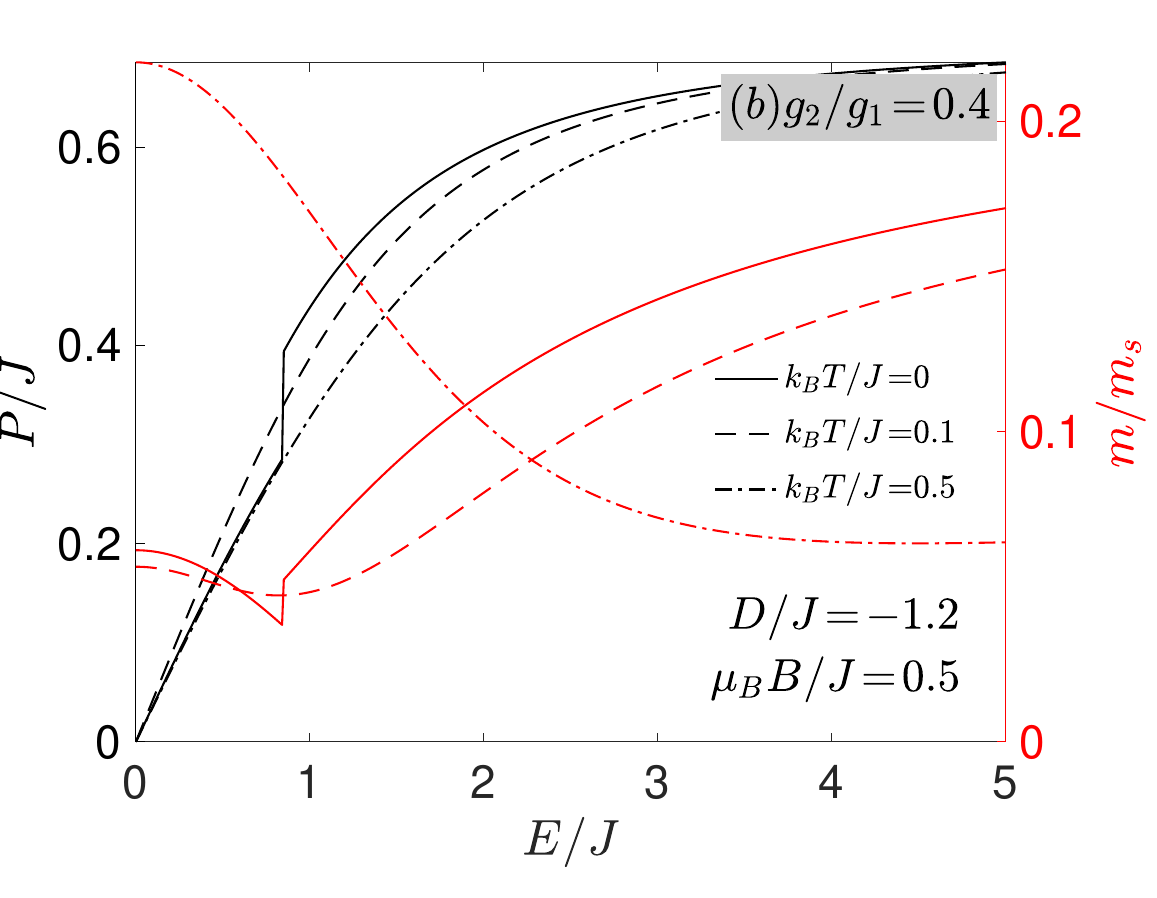}}{\includegraphics[width=0.33\textwidth,trim=0.9cm 7cm 0.5cm 7.5cm,, clip]{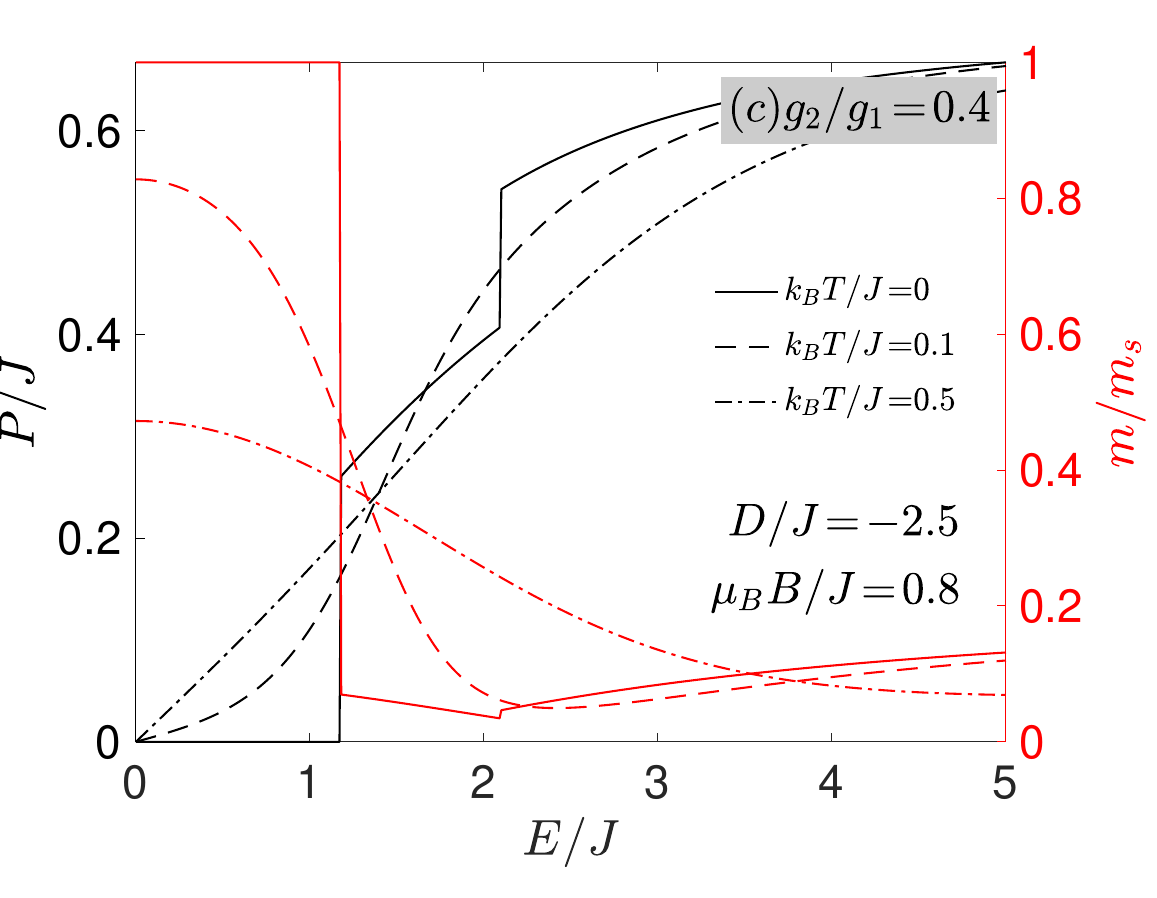}}
\caption{The dielectric polarization $P/J$ (left $y$-axis) and  normalized magnetization  $m/m_s$ (right $y$-axis) as  a function of the dimensionless electric  field energy $E/J$  calculated for three temperatures and selected set of  model parameters indicated in panels. }
\label{fig3}
\end{figure*}

\begin{figure*}[t!]
{\includegraphics[width=0.33\textwidth,trim=0.85cm 7cm 0.5cm 7.5cm,, clip]{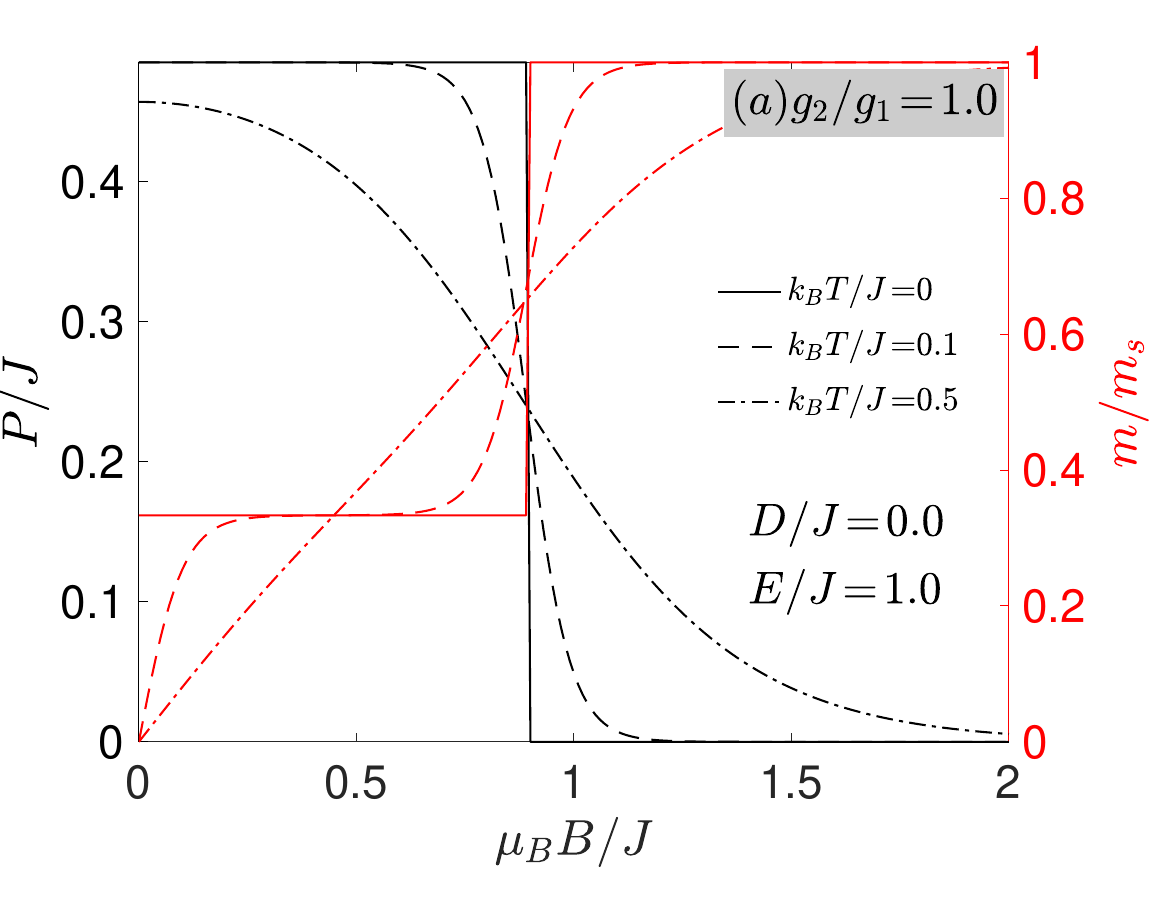}}{\includegraphics[width=0.33\textwidth,trim=0.75cm 7cm 0.5cm 7.5cm,, clip]{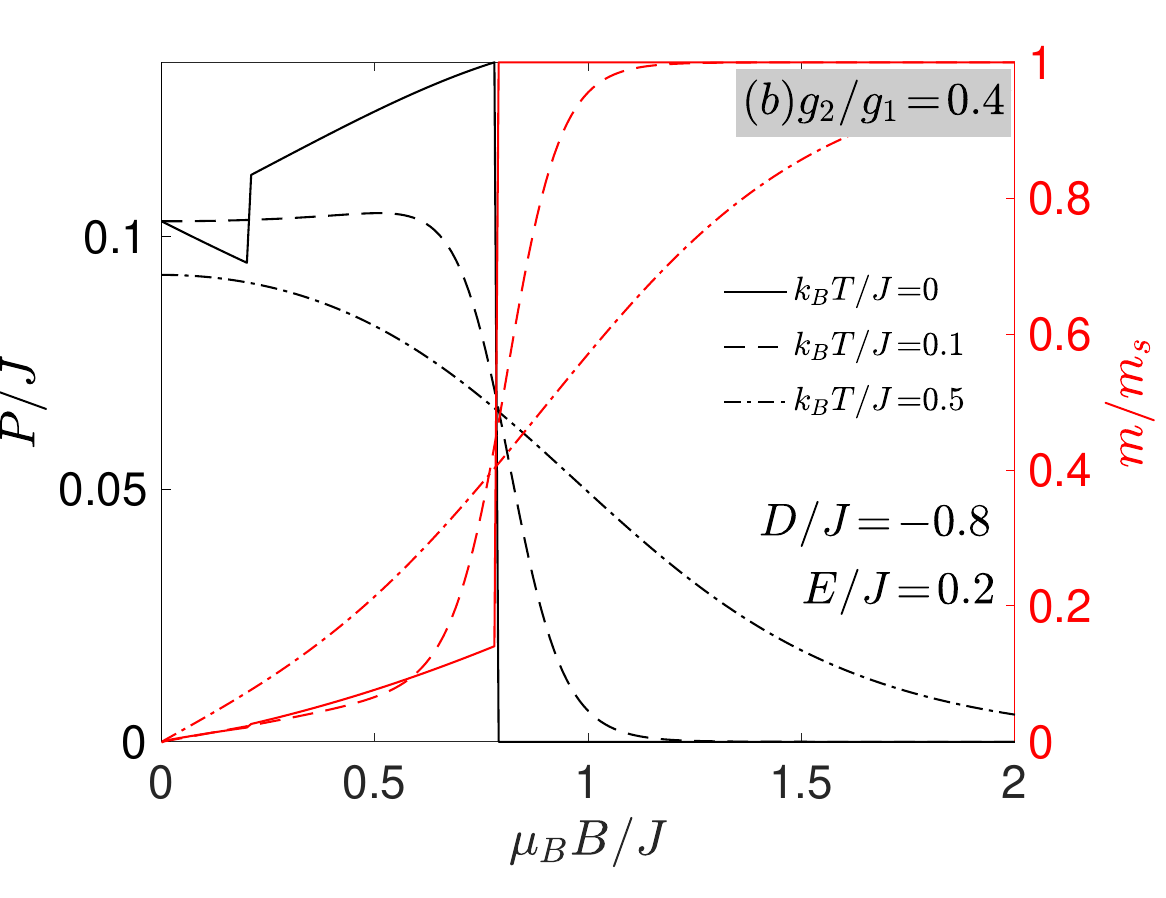}}{\includegraphics[width=0.33\textwidth,trim=0.85cm 7cm 0.5cm 7.5cm,, clip]{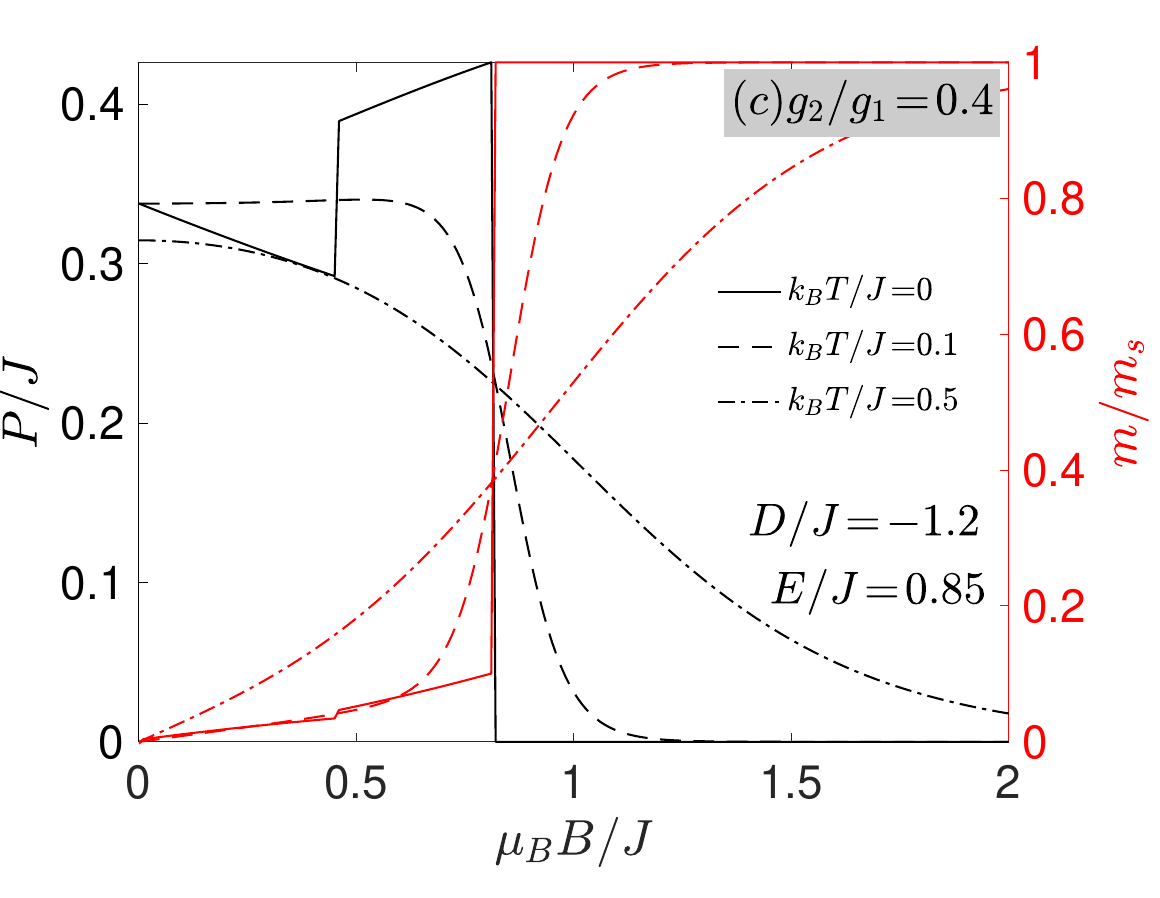}}
\caption{The dielectric polarization $P/J$ (left $y$-axis) and  normalized magnetization  $m/m_s$ (right $y$-axis) as a function of the  magnetic field $\mu_BB/J$ calculated for three temperatures and selected set of  model parameters indicated in panels. }
\label{fig4}
\end{figure*}

Now, let us look at the magnetoelectric effect of the mixed spin-(1/2, 1) Heisenberg dimer.  In the most general characteristic, the magnetoelectric effect could be determined through the magnetic-field dependence of the dielectric polarization and vice versa the electric-field dependence of the magnetization.  The relative size of dielectric polarization \eqref{eq13xa} is a direct consequence of the applied electric  and   magnetic fields, whereas the magnetic field by itself cannot induce the dielectric polarization in absence of the external electric field. Consequently, the 
spontaneous ferroelectricity  of the mixed spin-(1/2, 1) Heisenberg dimer is prohibited. On the other hand, the  ground-state magnetization normalized with respect to its saturation value $m/m_s$ in the quantum ferrimagnetic phases $\vert {\rm QF}_{\pm}\rangle$  is a monotonically decreasing  function of the  electric field energy $E$ even in absence of the external magnetic field
\begin{align}
\lim_{\frac{\mu_BB}{J}\to0}\langle {\rm QF}_{\pm}\vert \frac{m}{m_s}\vert {\rm QF}_{\pm}\rangle\!=\!\frac{1}{g_1\!+\!2g_2}\left[\pm g_2\!\mp\!\frac{(g_1\!-\!g_2)\left(J\!-\!2D\right)}{\sqrt{\left[(J\!-\!2D\right)^2\!+\!8\left[(J\Delta)^2\!+\!E^2\right]}} \right].
\label{eq22x}
\end{align}
Note furthermore that  arbitrarily small but non-zero thermal fluctuations completely destroy the spontaneous magnetization. The behavior of the magnetization and dielectric polarization are demonstrated in Figs.~\ref{fig3}-\ref{fig4} for a few selected sets of model parameters in presence of both external  fields. 
At zero temperature both quantities exhibit  stepwise changes in a vicinity of all field-driven phase transitions, which  gradually smear out upon increasing of temperature. The zero-temperature dependence of the dielectric polarization on the  electric field energy (Fig.~\ref{fig3}) demonstrate three different regimes conditioned by the ground-state spin arrangement.  The zero-temperature value of dielectric polarization  corresponding to the classical ferromagnetic phase $\vert {\rm F}_+\rangle$  is a constant function with respect to the electric field energy with a zero magnitude, \textit{i.e.} $P_{\rm F_+}/J\!=\!0$, whereas the dielectric  polarization  corresponding to the quantum ferrimagnetic phases $\vert {\rm QF}_{\mp}\rangle$  depends on the  electric field  and magnetic field thorough the following relations
\begin{align}
P_{\rm QF_{\pm}}\!=\!\frac{2E}{\sqrt{\left[J\!-\!2D\!\mp\!2(h_1\!-\!h_2)\right]^2\!+\!8[\left(J\Delta\right)^2\!+\!E^2]}}.
\label{eq23x}
\end{align}
In agreement with general expectations, the increasing  electric field enhances the dielectric polarization, while the  increasing magnetic field can reduce or enhance the dielectric polarization at  low enough temperatures as a result of different dielectric responses of the quantum ferrimagnetic ground states $\vert {\rm QF}_+\rangle$ and $\vert {\rm QF}_-\rangle$, respectively (Figs.~\ref{fig4}$(b)$ and $(c)$). It has been found, furthermore, that such extraordinary behavior can be detected if and only if $g_2/g_1\!\neq\!1$, because the identical values of Land\'e $g$-factors make the dielectric polarization \eqref{eq23x} fully independent of the magnetic field. 

Analyzing in detail the behavior of magnetization, one may reach the following interesting observations. (i) The ground-state magnetizations of the quantum ferrimagnetic phases $\vert {\rm QF}_{\pm}\rangle$  are in general complex  non-linear functions of the magnetic field
\begin{align}
\langle {\rm QF}_{\pm}\vert \frac{m}{m_s}\vert {\rm QF}_{\pm}\rangle\!=\!\frac{1}{g_1\!+\!2g_2}\left[\pm g_2\!\mp\!\frac{(g_1\!-\!g_2)\left[J\!-\!2D\!\mp\!2(h_1\!-\!h_2)\right]}{\sqrt{\left[(J\!-\!2D\!\mp\!2(h_1\!-\!h_2)\right]^2\!+\!8\left[(J\Delta)^2\!+\!E^2\right]}} \right].
\label{eq24x}
\end{align}
 (ii) The zero-temperature magnetization   exhibits a significant local minimum as a function of the electric field energy at its specific value, which appears in the vicinity of the  electric-field driven phase transition between both quantum ferrimagnetic ground states $\vert {\rm QF}_{\pm}\rangle$ [Fig.~\ref{fig3}$(b)$ and $(c)$]. (iii) The magnetization of the quantum ferrimagnetic $\vert {\rm QF}_+\rangle$ phase saturating  at $m/m_s\!=\!1/3$ for the  fully isotropic case $g_2/g_1\!=\!1$ changes its curvature as well as its position when considering the anisotropic case  $g_2/g_1\!\neq\!1$  [see Figs.~\ref{fig4}$(b)$ and $(c)$].  Owing to this fact,  the  magnetization quasi-plateau can be observed at moderate magnetic fields instead of the true one-third magnetization plateau. The   magnetization quasi-plateau connected to the $\vert {\rm QF}_+\rangle$ phase is positioned below the value $m/m_s\!=\!1/3$ for $g_2/g_1\!<\!1$, whereas the   magnetization quasi-plateau for $g_2/g_1\!>\!1$  is always detected above the value  $m/m_s\!>\!1/3$.  Besides, the non-zero electric field stabilizes  existence of quasi-plateaus in response to  the magnetic field and/or temperature, because the increasing electric field energy has tendency to align  magnetic moments in a perpendicular direction  with respect to the applied  magnetic field. The ground-state magnetization corresponding to the other quantum ferrimagnetic $\vert {\rm QF}_-\rangle$ phase  again leads to the quasi-plateau, which   converges to $m/m_s\!=\!0$ as one approaches the zero magnetic field. It can be  deduced from Figs.~\ref{fig3} and \ref{fig4} that  the striking magnetization quasi-plateau  is perfectly mirrored also in a quasi-plateau behavior of the dielectric polarization, which markedly demonstrates the magnetoelectric  coupling in zero as well as non-zero temperatures of the mixed spin-(1/2, 1) Heisenberg dimer.

\subsection{Magnetocaloric and electrocaloric properties}

\begin{figure*}[b!]
{\includegraphics[width=0.33\textwidth,trim=1cm 8cm 0.5cm 7.5cm,, clip]{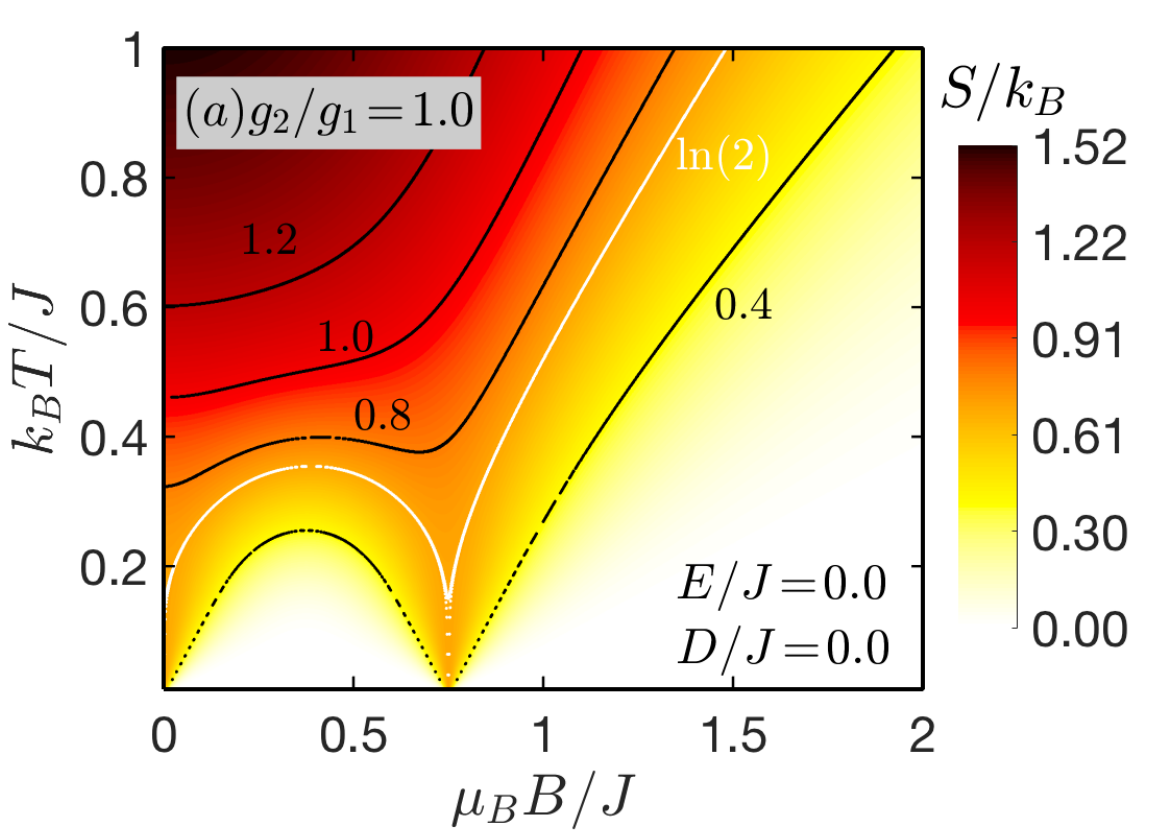}}
{\includegraphics[width=0.33\textwidth,trim=1cm 8cm 0.5cm 7.5cm, clip]{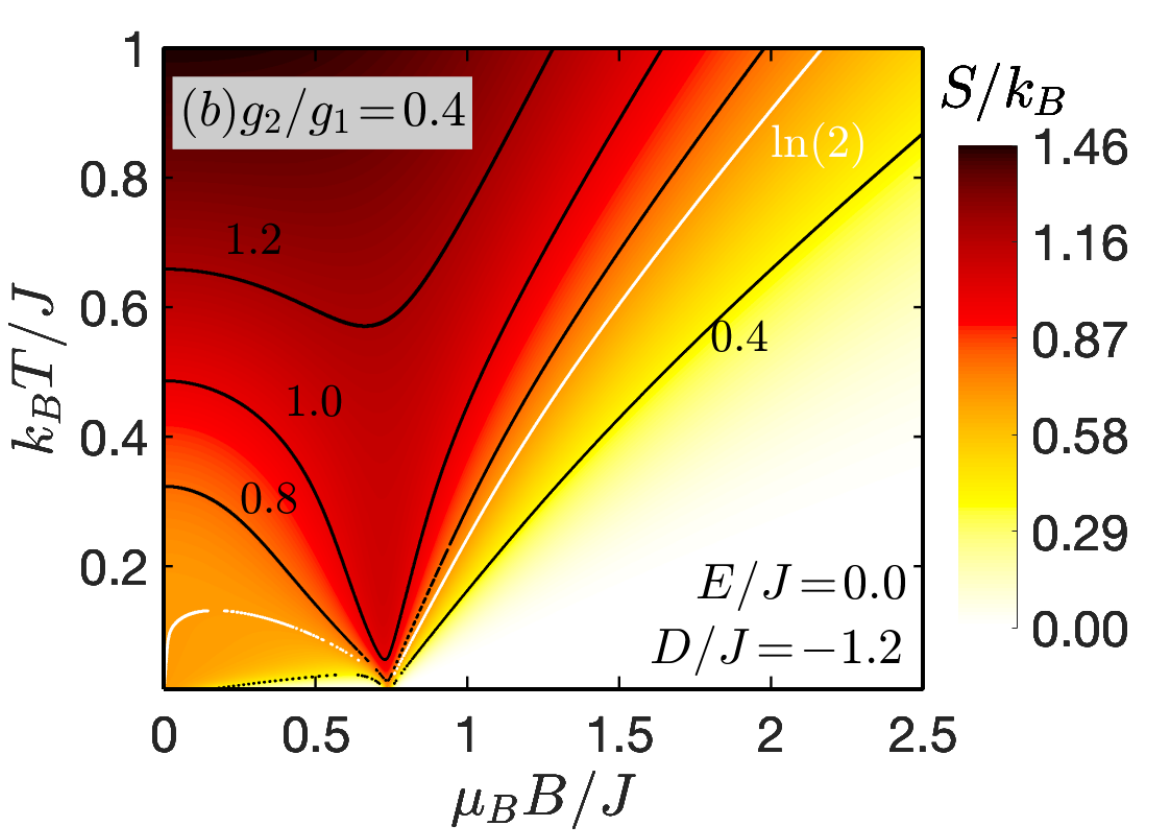}}
{\includegraphics[width=0.33\textwidth,trim=1cm 8cm 0.5cm 7.5cm, clip]{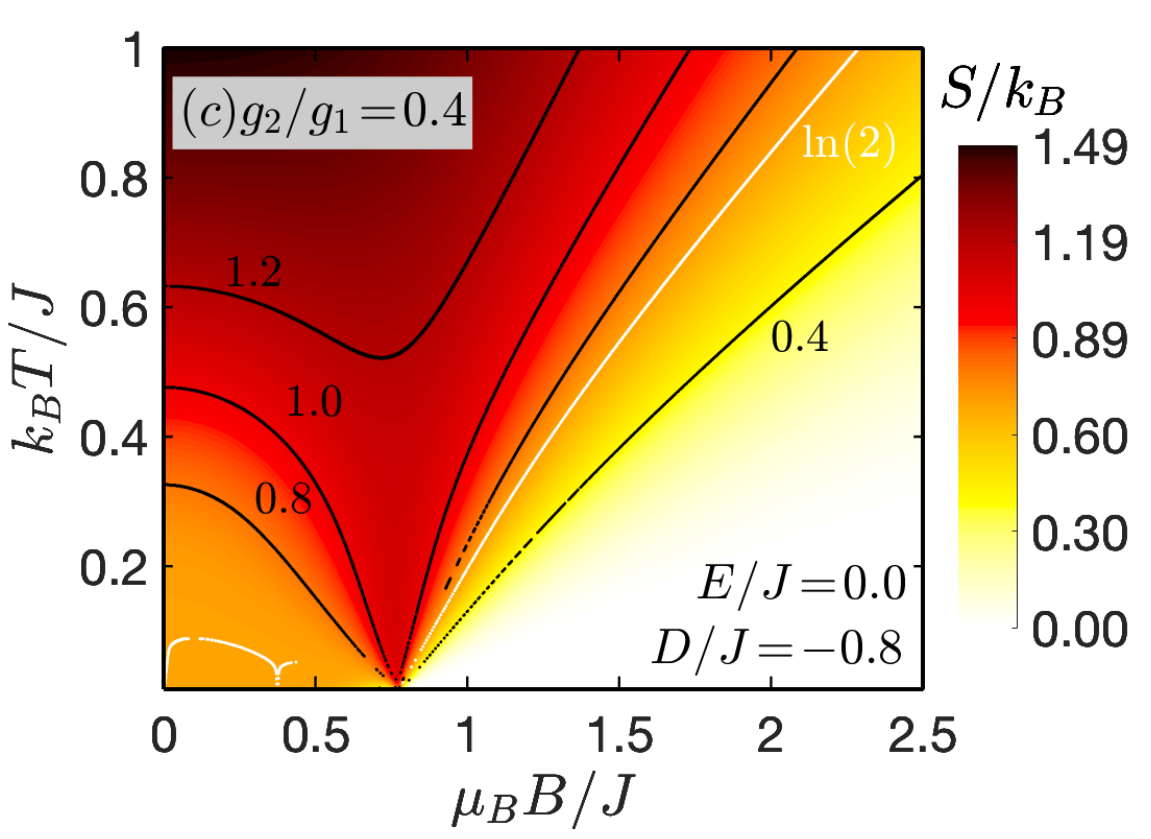}
}\caption{Density plots of the entropy in the     magnetic field  versus temperature plane for three selected sets of the model parameters in absence of an applied  electric field. Displayed curves in all  panels demonstrate isentropic lines, among which  particular isentropy lines with $S/k_B\!=\!\ln(2)$ are highlighted by a white color. }
\label{fig5}
\end{figure*}

In order to study the magnetocaloric and electrocaloric properties of a mixed spin-(1/2, 1) Heisenberg dimer  we will analyze in detail the adiabatic changes of  temperature achieved upon  variation of  the external magnetic  or electric field.   It should be expected that in the proximity of a certain field-driven phase transition, the competition between all driving forces can lead to the more or less rapid change of temperature, however the system entropy remains fixed. The slope as well as the distance of two neighboring temperatures are two crucial identifiers of intensity of cooling process. The density plots of the entropy  as a function of temperature and  magnetic field are presented in Fig.~\ref{fig5}. For isentropic changes of  temperature driven exclusively by an applied magnetic field  ($E/J\!=\!0.0$) we identify three different scenarios. 
The density plot of the entropy for  high enough ratio $g_2/g_1$ between Land\'e $g$-factors,  for which only the $\vert {\rm F}_+\rangle$ and  $\vert {\rm QF}_{+}\rangle$  ground states are available, is   presented in Fig.~\ref{fig5}$(a)$.  It is quite apparent that the most prominent adiabatic change of temperature is observed close to the value of the residual entropy $S/k_B\!=\!\ln(2)$, along which the temperature drops infinitely fast in the vicinity of  the  transition field between  $\vert {\rm QF}_{+}\rangle$-$\vert {\rm F}_+\rangle$  phases. 
Almost identical  temperature change is  identified  very close to   zero magnetic field, where  two-fold degeneracy is lifted by   an arbitrarily small non-zero magnetic field.  Because  the single-ion anisotropy $D/J$ reduces the effect of magnetic field, its increasing magnitude can be used as an additional tuning parameter for enlargement of the temperature  interval with an infinite change of temperature.
It was identified furthermore, that the difference of both Land\'e $g$-factors is other crucial tuning parameter, leading to the  imbalance between effectiveness of MCE (temperature span of infinite change) in zero and non-zero transition fields. The other particular case is shown in Fig.~\ref{fig5}$(b)$, where  the  most rapid change of temperature is again identified at a zero magnetic field  along the isentropic line  $S/k_B\!=\!\ln(2)$. Contrary to the previous case the magnetocaloric effect in a vicinity of the finite transition field, corresponds to  a zero-temperature phase transition between   $\vert {\rm QF}_{-}\rangle$ and $\vert {\rm F}_+\rangle$ phases is more significant along the isentropic line slightly below the $S/k_B\!=\!\ln(3)$, nevertheless the most rapid changes are realized at very narrow temperature interval shifted to  higher temperatures. 
The  third  scenario, which is illustrated in  Fig.~\ref{fig5}$(c)$, involves two field-driven discontinuous phase transitions $\vert {\rm QF}_{-}\rangle$-$\vert {\rm QF}_{+}\rangle$-$\vert {\rm F}_{+}\rangle$. The most pronounced change of temperature is repeatedly detected along the isentropic line calculated for the   entropy $S/k_B\!=\!\ln(2)$, which involves three visible  minimas at zero field and in a proximity of  transition fields associated with the field-induced transitions  $\vert {\rm QF}_{-}\rangle$-$\vert {\rm QF}_{+}\rangle$  and $\vert {\rm QF}_{+}\rangle$-$\vert {\rm F}_{+}\rangle$, respectively.   For the selected set of model parameters presented in Fig.~\ref{fig5}$(c)$ the temperature span of a rapid cooling process decreases with increasing magnetic transition point.

\begin{figure*}[t!]
{\includegraphics[width=0.33\textwidth,trim=1cm 8cm 0.5cm 7.5cm, clip]{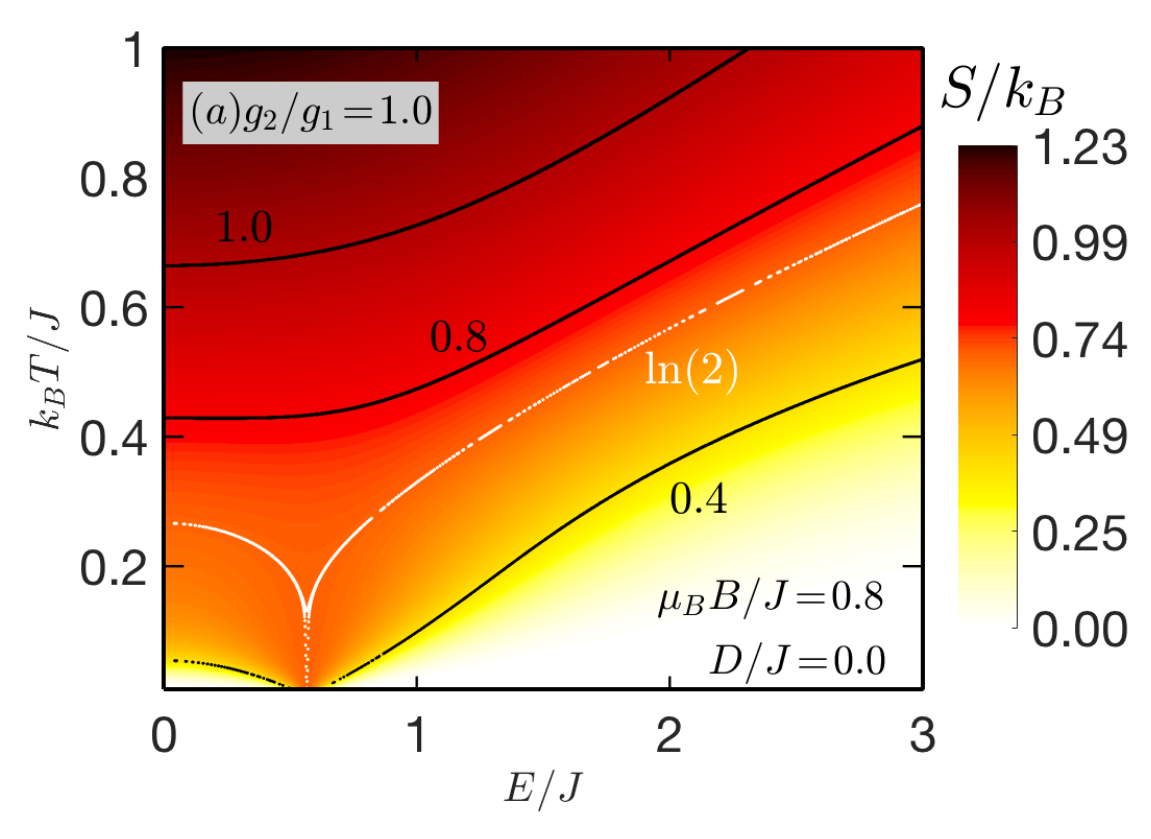}}
{\includegraphics[width=0.33\textwidth,trim=1cm 8cm 0.5cm 7.5cm, clip]{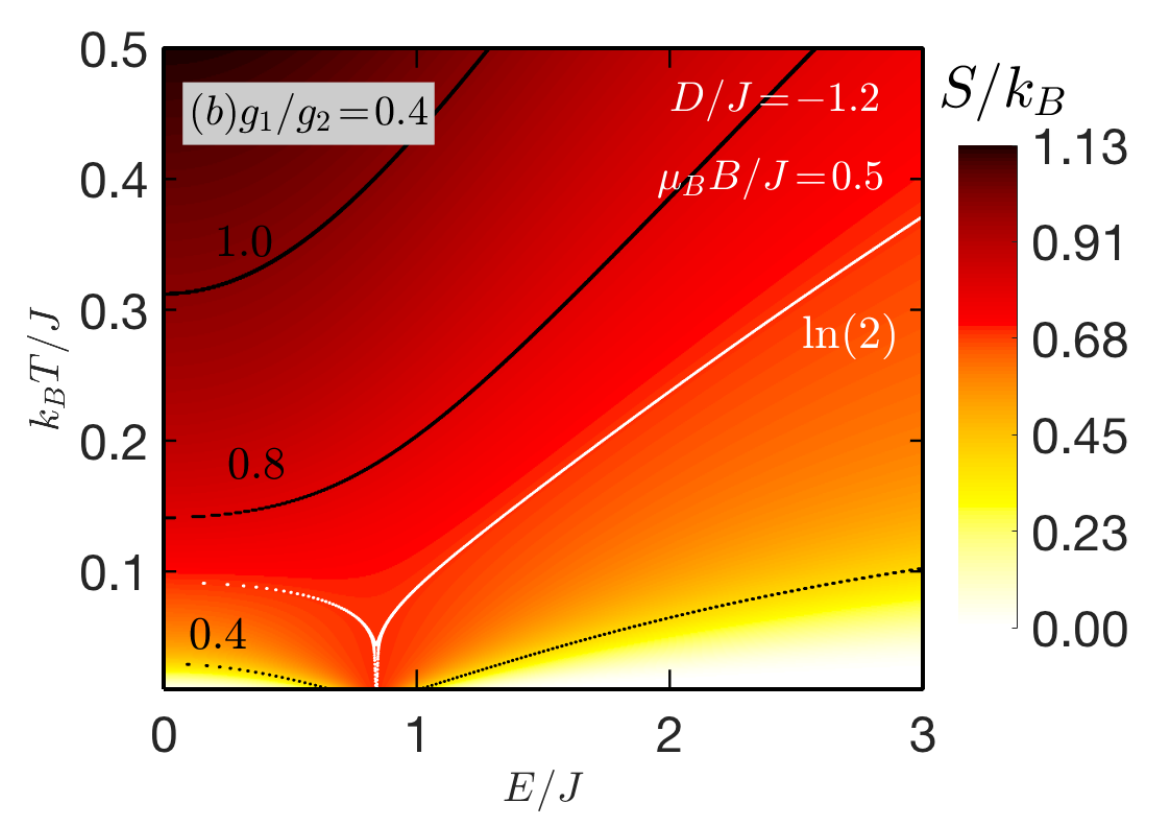}}
{\includegraphics[width=0.33\textwidth,trim=1cm 8cm 0.5cm 7.5cm, clip]{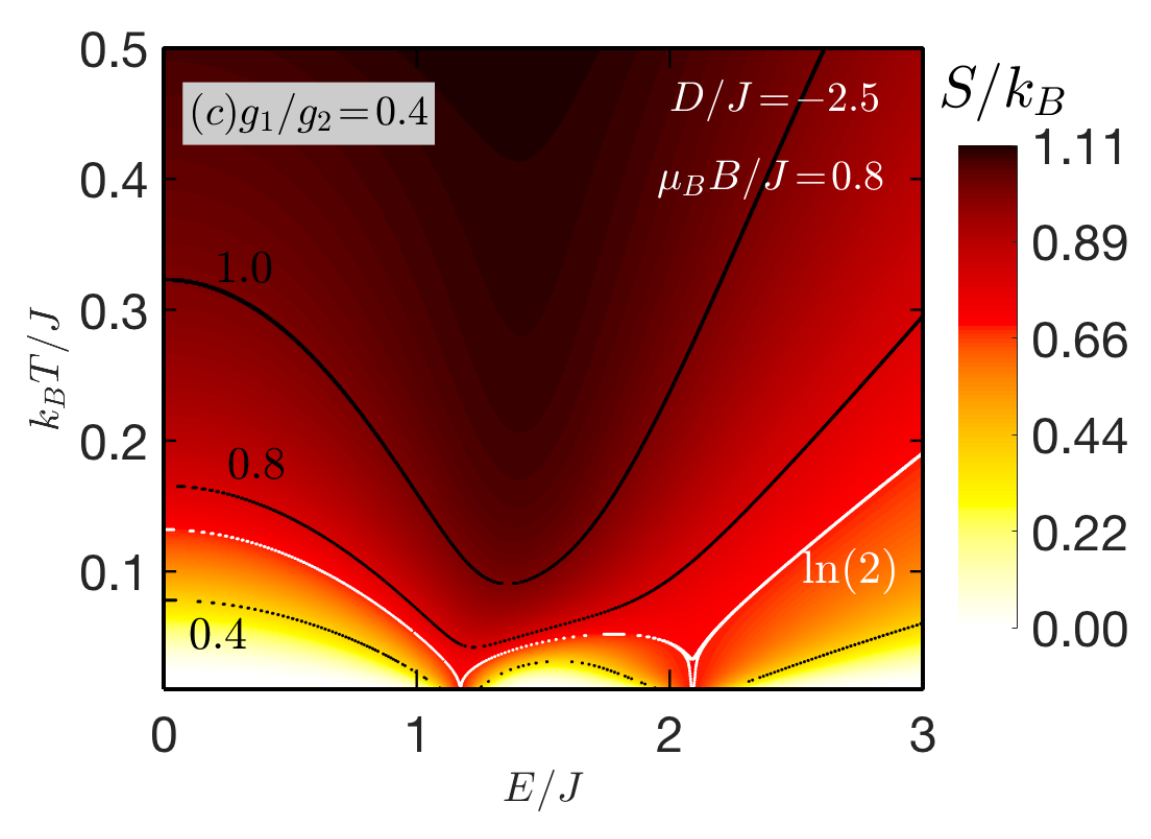}}
\caption{Density plots of the entropy in the   electric  field energy versus temperature plane for three selected sets of the model parameters in presence of an applied magnetic field. Displayed curves in all  panels show isentropic lines, among which a particular isoline with $S/k_B\!=\!\ln(2)$ is highlighted by a white color. }
\label{fig6}
\end{figure*}

The most  interesting observations follow from Fig.~\ref{fig6},  where the density plot of entropy is shown in the electric field energy versus temperature plane.
 In the range of magnetic fields, where the additional electric field  leads to the electric-field driven phase transition (see Fig.~\ref{fig2}) one can identify prolative infinitely fast change of temperature along the isoentropic line with $S/k_B\!=\!\ln(2)$. Consequently, the application of an additional electric field energy can enhance an observed MCE. Similarly as in the previously discussed MCE, the cooling temperature span can be tuned by variation of the strength of the single-ion anisotropy $D/J$ and/or the ratio of the Land\'e $g$-factors. Interestingly, there exists a parametric space, characterized through the existence of a triple point, where the increasing electric field  can stimulate the cooling procedure with two consecutive cooling points located in a vicinity of the electric-field driven $\vert {\rm F}_{+}\rangle$-$\vert {\rm QF}_{-}\rangle$ and $\vert {\rm QF}_{-}\rangle$-$\vert {\rm QF}_{+}\rangle$ phase transitions (for an illustration see Fig.~\ref{fig6}$(c)$).  Based on the analysis of the ground-state phase diagram (Fig.~\ref{fig2}) this parameter space is strongly conditioned by the strength of the single-ion anisotropy $D/J$, magnetic field $\mu_BB/J$ as well as the ratio $g_2/g_1$.

\begin{figure*}[b!]
{\includegraphics[width=0.33\textwidth,trim=1cm 8cm 0.5cm 7.5cm, clip]{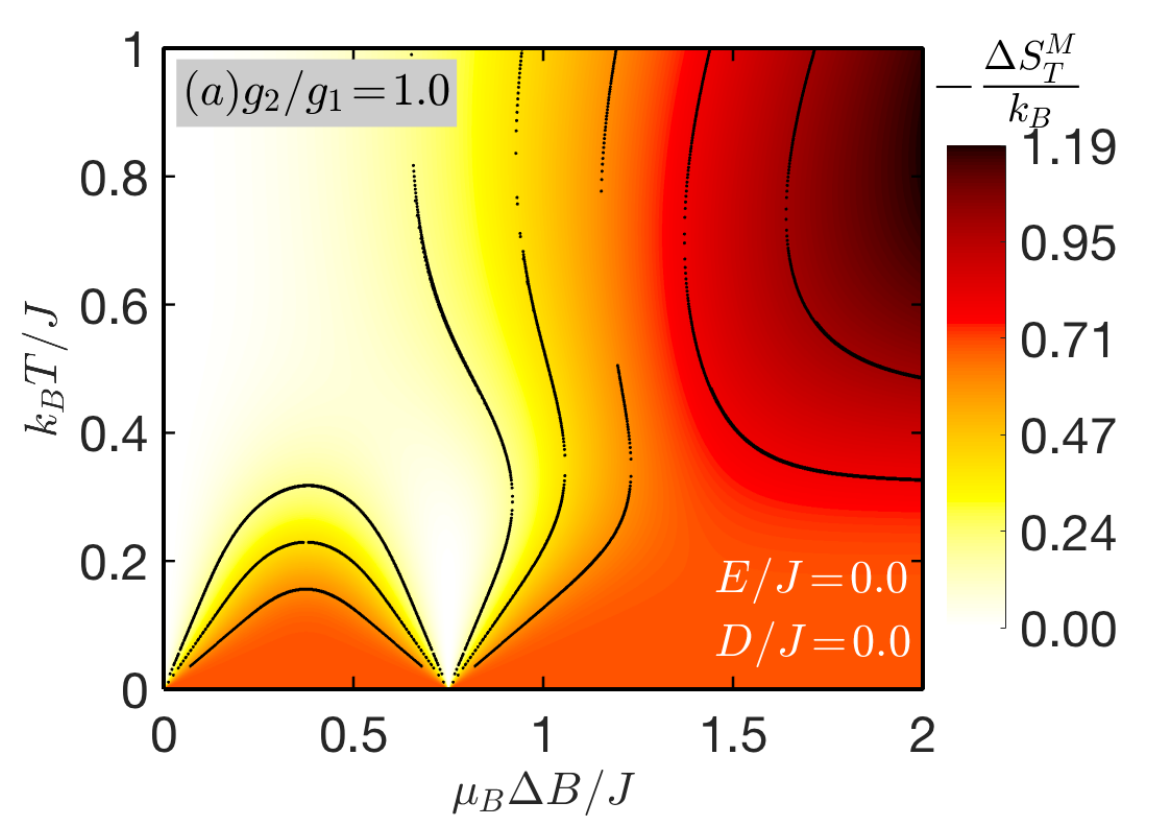}}
{\includegraphics[width=0.33\textwidth,trim=1cm 8cm 0.5cm 7.5cm, clip]{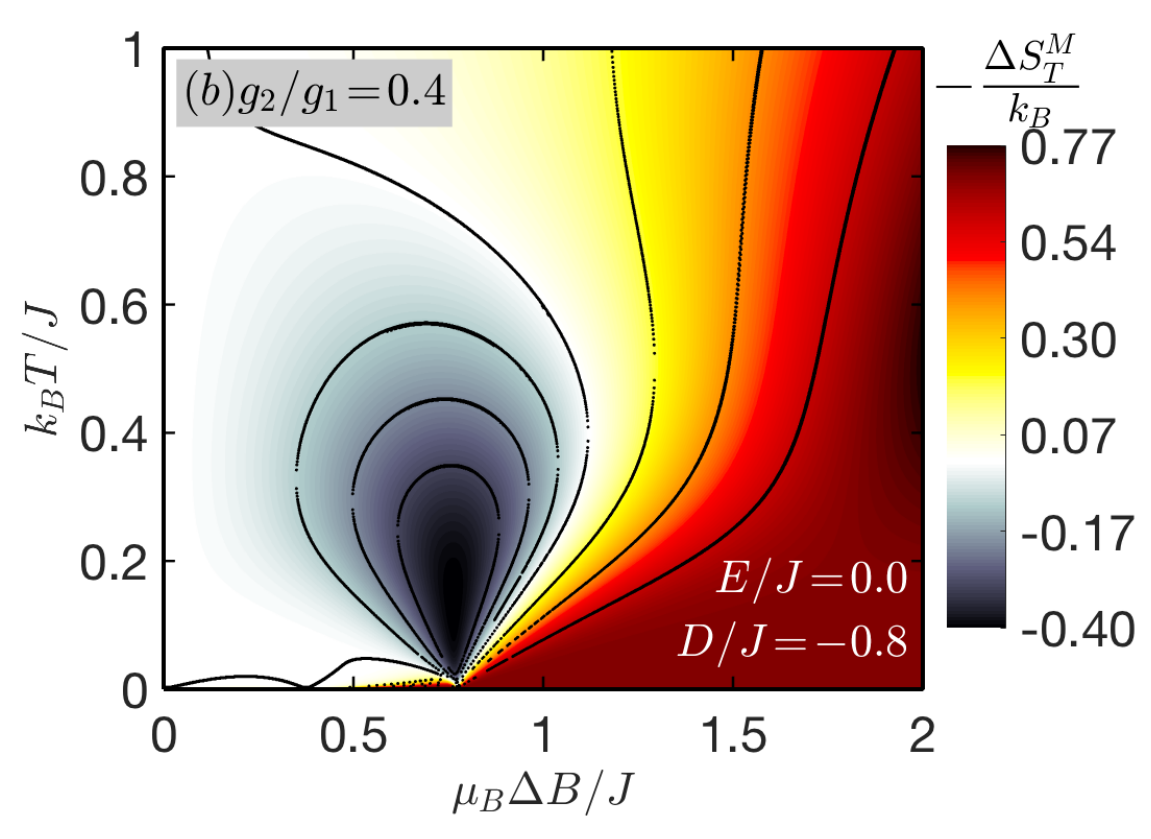}}
{\includegraphics[width=0.33\textwidth,trim=1cm 8cm 0.5cm 7.5cm, clip]{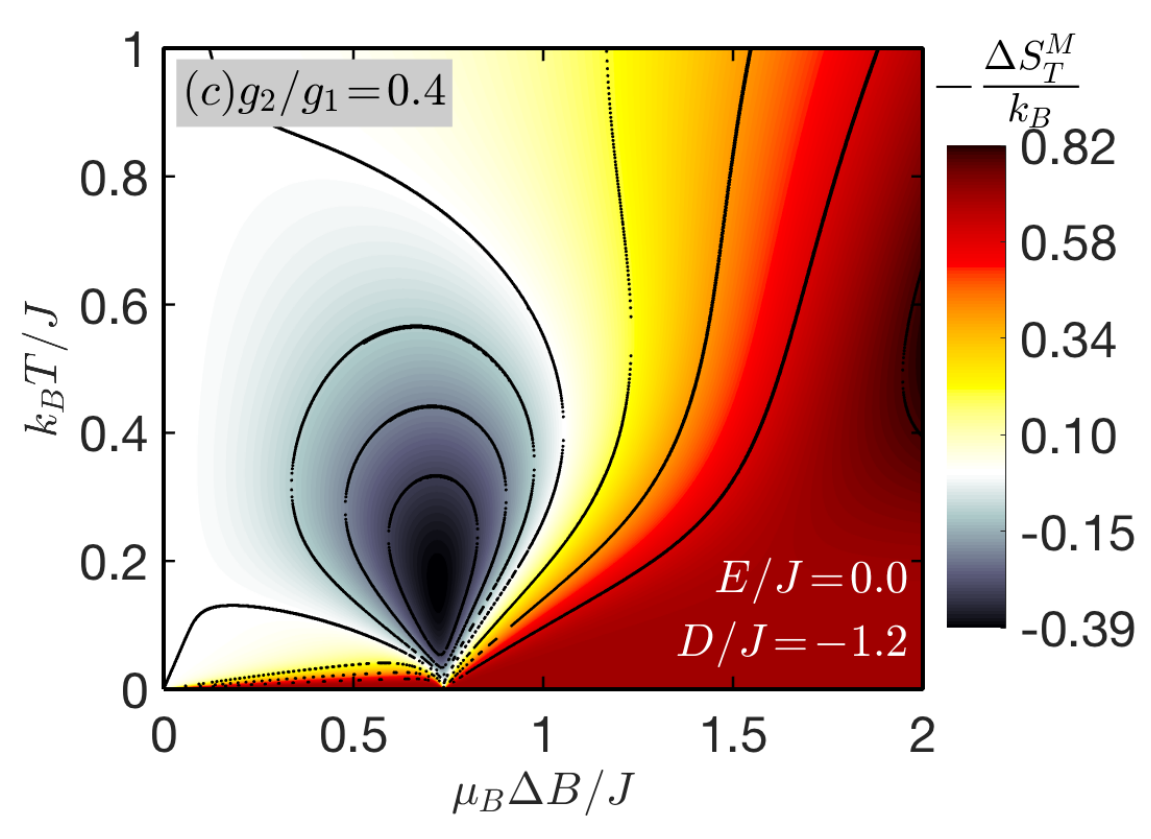}}
\caption{Density plots of the isothermal entropy change in the  magnetic-field change versus  temperature plane for three selected sets of  model parameters in absence of an applied electric field. Displayed curves in all  panels show isovalue lines running from $-\Delta S_T^M/k_B\!=\!-0.3$ to 1.2 with the step 0.1 (0.2) in a negative (positive) range. }
\label{fig7}
\end{figure*}

 From the experimental point of view it is more advisable to analyze 
the isothermal entropy change $\Delta S_{T}^M$,  an indirectly measured experimental quantity, defined as  a difference of a magnetic entropy at non-zero and zero magnetic field  at a fixed temperature 
\begin{align}
\Delta S_{T}^M(T, \mu_B\Delta B)&\!=\!S_2(T,\mu_BB\!\neq\! 0)-S_1(T,\mu_BB\!=\!0).
\label{eq25x}
\end{align}
In such definition, the maximal magnitude of the isothermal entropy change $|\Delta S_T^M|$ corresponds to the maximal intensity of the MCE driven by an external field at a fixed temperature.  It should be noted that in general, the  isothermal entropy change $\Delta S_{T}^M$ can achieve both the positive as well as negative values, which characterize two different types of MCE.  In the present convention  $-\Delta S_T^M\!>\!0$  corresponds to the conventional MCE, whereas $-\Delta S_T^M\!<\!0$ is  characteristic feature of  a more exotic inverse MCE.
The exact results obtained for the same model parameters as in Fig.~\ref{fig5} are presented in Fig.~\ref{fig7}.
 For the fully isotropic case (Fig.~\ref{fig7}$(a)$) only the conventional MCE is predicted for the whole range of magnetic field and temperature. The global maximum, and thus the most pronounced MCE,  occurs approximately at relatively high temperature $k_BT/J\!\approx \!0.8$ at magnetic field where the $\vert {\rm  F}_+\rangle$ ground state is favored. In the region of $\vert {\rm  QF}_+\rangle$ ground state the maximal isothermal entropy change is gradually destroyed by thermal fluctuations. Naturally,  the maximal thermal stability should be expected at the  center of both transition points, where the impact of neighboring phases is minimal. 
Contrary to this,  Figs.~\ref{fig7}$(b)$ and~\ref{fig7}$(c)$ indicate existence of  both the conventional as well as inverse MCE in a certain region of the parameter space.  The maximum of the conventional MCE is again detected at high-enough magnetic field but at significantly smaller temperature in comparison to the isotropic case.   The inverse MCE is predicted at the proximity of the electric-field driven phase transition with a relatively low  thermal stability depending on remaining model parameters. It should be emphasized, that the inverse MCE can arise  in case of imbalanced Land\'e $g$-factors only.

\begin{figure*}[t!]
\centering
{\includegraphics[width=0.32\textwidth,trim=0.5cm 7.5cm 2cm 8cm, clip]{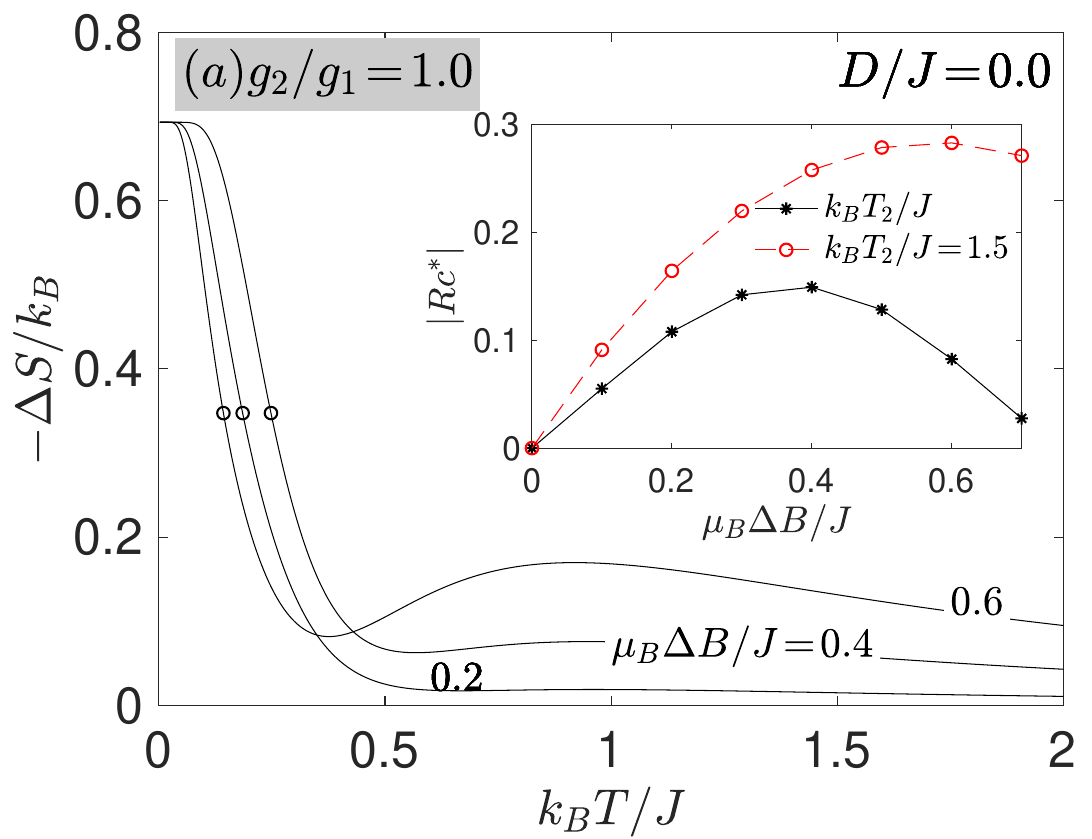}}
{\includegraphics[width=0.32\textwidth,trim=0.5cm 7.5cm 2cm 8cm, clip]{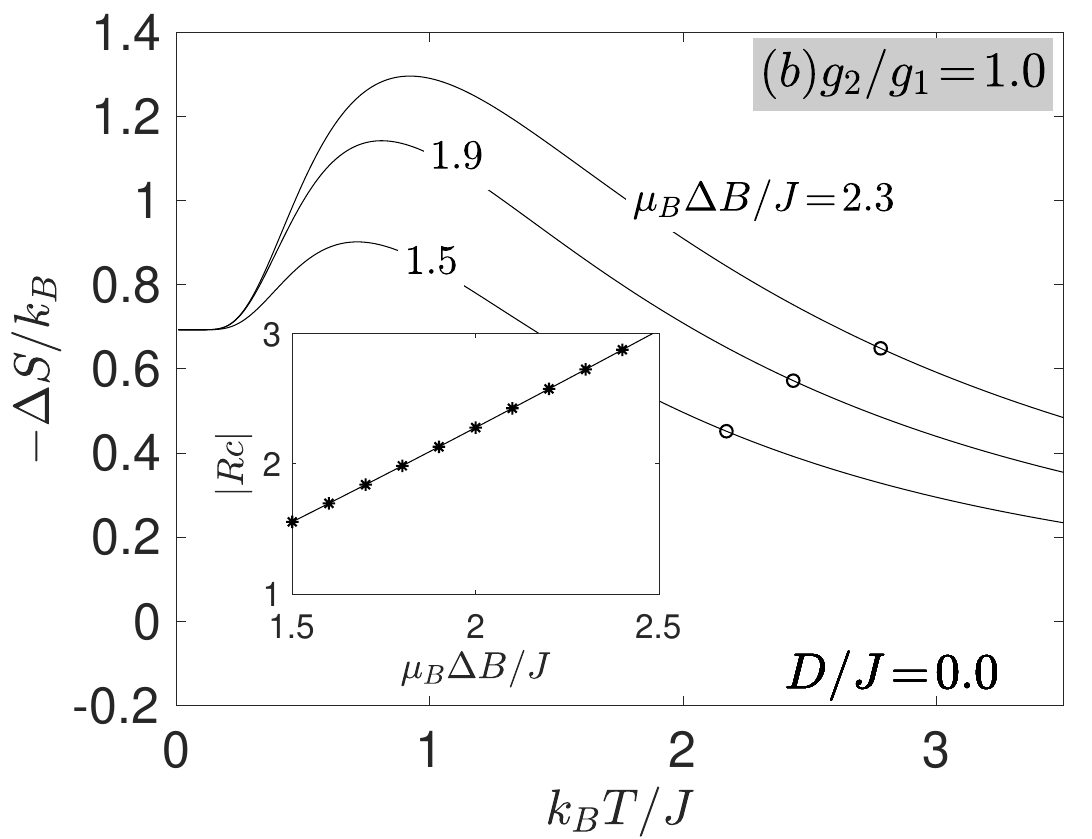}}
{\includegraphics[width=0.32\textwidth,trim=0.5cm 7.5cm 2cm 8cm, clip]{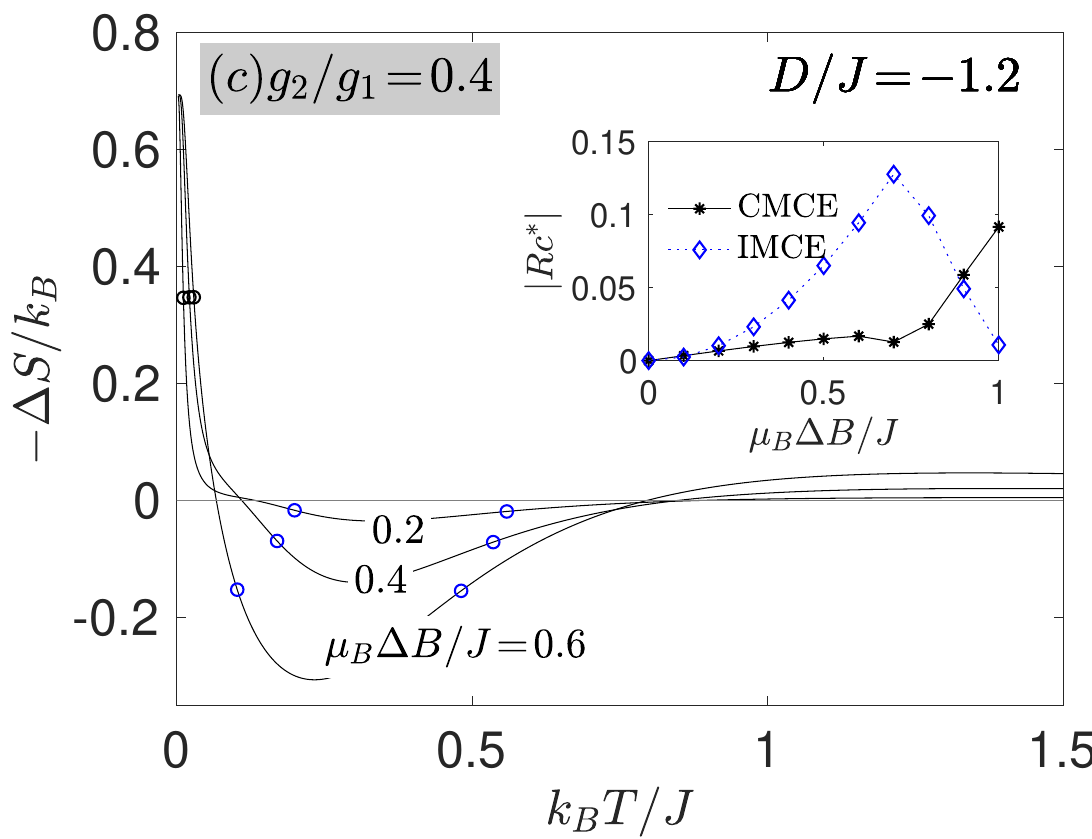}}
\caption{The dependence of a negative isothermal entropy change on the temperature for selected values of $\mu_B\Delta B/J$ and selected set of $D/J$, $g_2/g_1$ parameters. Open black and blue circles  in main panels denote limiting temperatures emerging in the Eq.~\eqref{eq25xa} to calculate the $|Rc|$ coefficient for the  conventional MCE (CMCE) and inverse MCE (IMCE), respectively. Insets: The evolution of the $|Rc|$ coefficient under the external magnetic field span. Black (blue) curves illustrate the change of $Rc$ between $T_1$ and $T_2$ temperature for the conventional (inverse) MCE, while the red curve in the inset of panel $(a)$ illustrates the enlargement of $|Rc|$ if the upper fixed temperature limit, $k_BT_2/J\!=\!1.5$.}
\label{fig7x}
\end{figure*}

Now let us look on the cross-section of the Fig.~\ref{fig7} to gain more insight into the behavior of the isothermal entropy change. The plots Fig.~\ref{fig7x}$(a)$ and $(b)$  illustrate the influence of the magnetic field on the isothermal entropy change $-\Delta S/k_B$ with an absence of electric field for fully isotropic case together with the other relevant quantity, the refrigerant capacity~\cite{Pecharsky}
\begin{align}
Rc=\int_{T_1}^{T_2}\Delta S_{T}^MdT,
\label{eq25xa}
\end{align}
which allows us to calculate the performance of a refrigeration cycle and thus to quantify the amount of heat transferred between hot and cold reservoirs at temperatures  $T_1$ and $T_2$. The limiting temperatures are selected as an intersection of integrated function and the constant function parallel to the $x$-axis passing through the half of maximal amplitude.
\\
For the high-enough magnetic field, where the spontaneous $\vert {\rm F}_+\rangle$ phase is preferred (Fig.~\ref{fig7x}$(b)$) the isothermal entropy change is gradually enlarged with an increasing magnetic field. The magnitude as well as the position of maximum are also gradually enlarged. The refrigerant capacity (inset in Fig.~\ref{fig7x}$(b)$) clearly demonstrates  an enhancement of the MCE driven by a magnetic field, which exhibits an expected linear character typical for the ferromagnetic phase. It should be emphasized, that the qualitatively identical behavior is detected for an arbitrary ratio $g_2/g_1$, if the $\vert {\rm F}_+\rangle$ ground state is realized.  In the low-enough magnetic field, connected to the spontaneous $\vert {\rm QF}_+\rangle$ phase (Fig.~\ref{fig7x}$(a)$) the magnitude as well as the position of low-temperature maximum remains unchanged, while the width of maximum is varied upon the variation of magnetic field. The maximal isothermal entropy change is observed around the magnetic-field change $\mu_B\Delta B/J\!\sim\!0.4$ (black curve in  the inset of Fig.~\ref{fig7x}$(a)$).  Towards  the one of the transition magnetic field, the interplay between the thermal fluctuations and the energy equivalence of both neighboring phases declines the heat transfer and thus the MCE decreases. Focusing on  the higher (fixed) temperature interval (red curve in the inset of Fig.~\ref{fig7x}$(a)$) the refrigerant capacity can shift its maximum to the higher magnetic field as a consequence of the second broad maximum in the isothermal entropy change. The existence of an inverse MCE at finite temperatures  gives rise to  other extreme with an opposite sign. As illustrates Fig.~\ref{fig7x}$(c)$ the increasing magnetic field firstly enhances the magnitude of the isothermal entropy change reaching the maximum around the transition magnetic field.  Subsequently, the negative minimum in isothermal entropy change vanishes similarly as the refrigerant capacity (blue dotted curve in the inset of Fig.~\ref{fig7x}$(c)$).  On the other hand, the refrigerant capacity of the conventional MCE is minimal in the same region of $\vert {\rm QF}_-\rangle$ ground state, but rapidly enlarges when the transition point to the $\vert {\rm F}_+\rangle$ state is reached.

\begin{figure*}[t!]
{\includegraphics[width=0.33\textwidth,trim=1cm 8cm 0.5cm 7.5cm, clip]{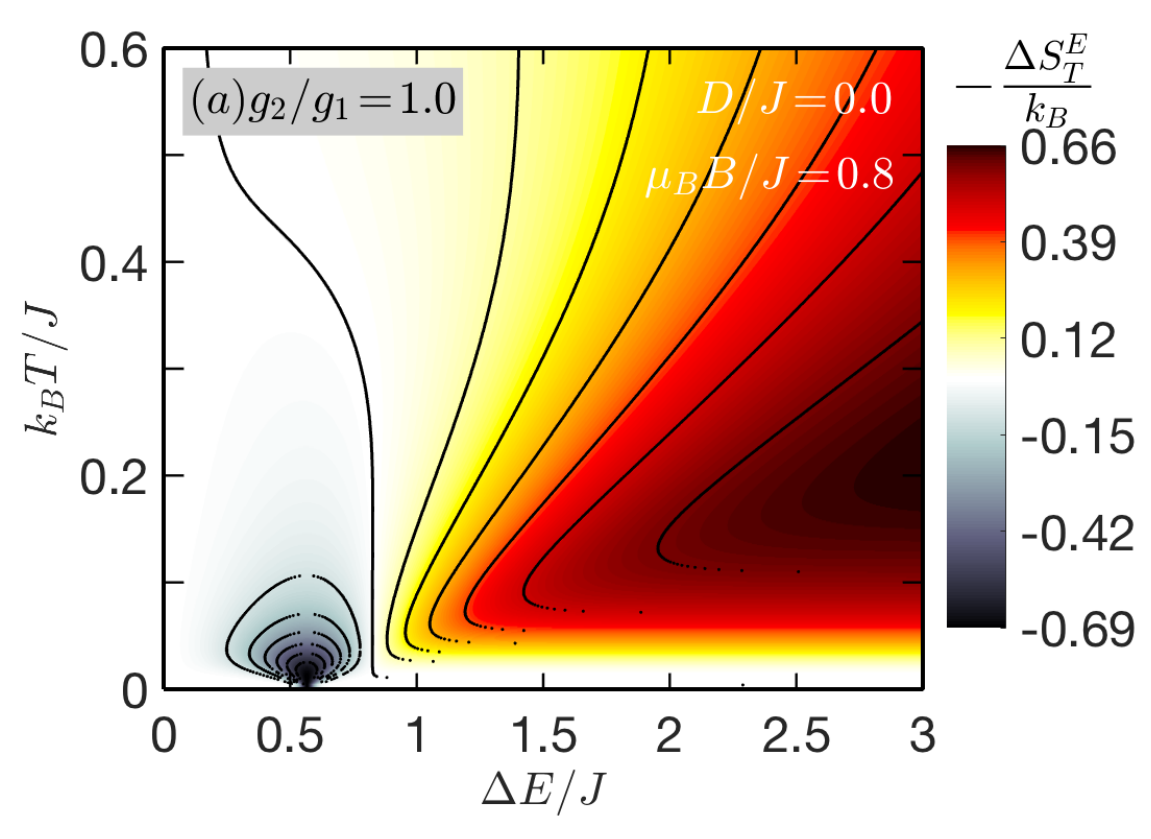}}
{\includegraphics[width=0.33\textwidth,trim=1cm 8cm 0.5cm 7.5cm, clip]{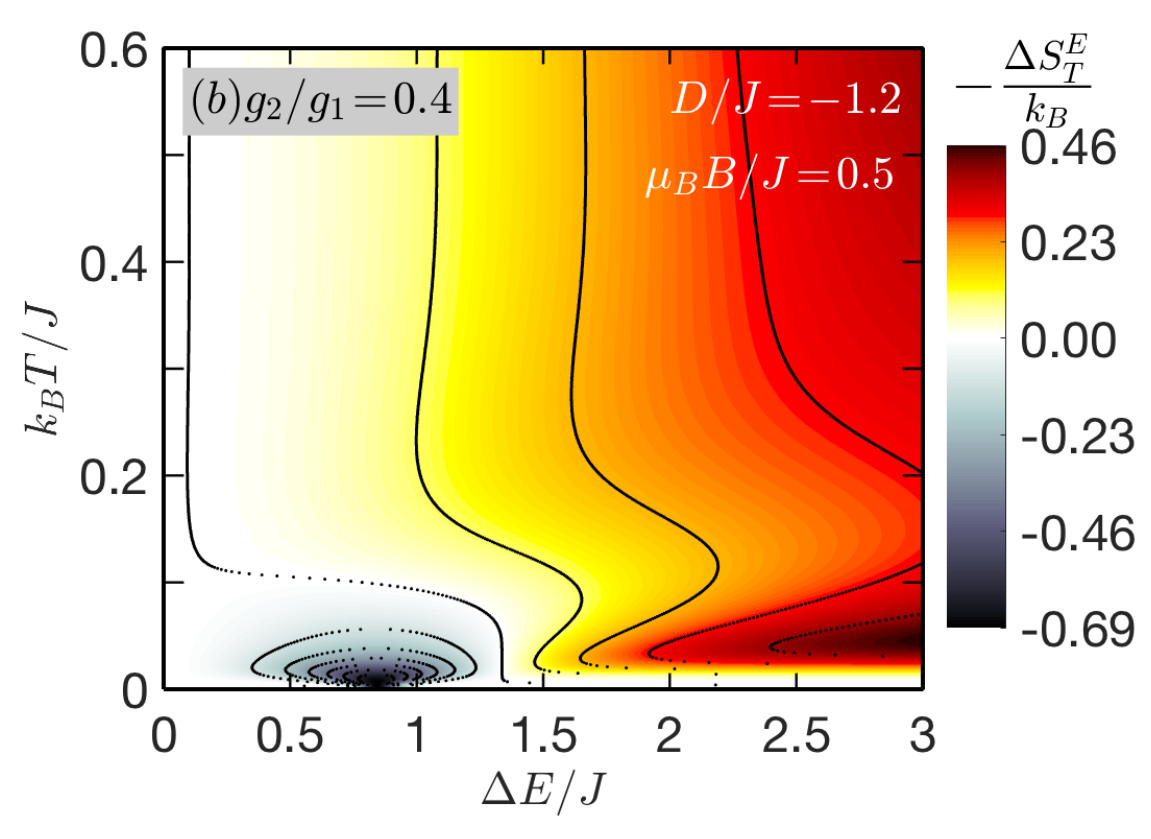}}
{\includegraphics[width=0.33\textwidth,trim=1cm 8cm 0.5cm 7.5cm, clip]{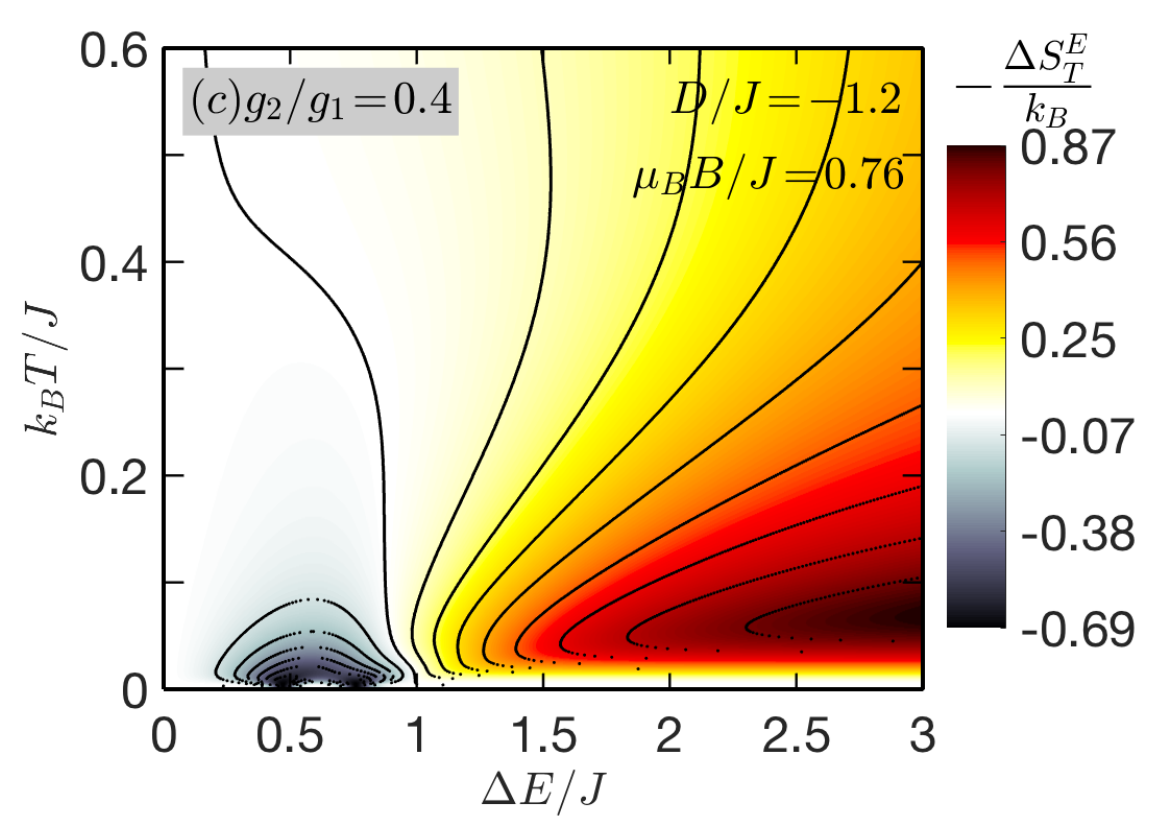}}
\caption{Density plots of the isothermal entropy change in   the electric-field-change  versus temperature plane for three selected sets of  model parameters in presence of an applied magnetic field. Displayed curves in all  panels show isovalue lines running from $-\Delta S_T^E/k_B\!=\!-0.7$ to 0.7 with the step 0.1. }
\label{fig8x}
\end{figure*}

In an analogy to the isothermal entropy changes induced by the magnetic field we have similarly examined the isothermal entropy changes induced by the electric field energy characterizing the ECE at a certain magnetic field 
\begin{align}
\Delta S^E_{T}(T,\mu_BB,\Delta E)&\!=\!S_2(T,\mu_BB,E\!\neq\! 0)-S_1(T,\mu_BB,E\!=\!0).
\label{eq26x}
\end{align}
A few illustrative examples of the isothermal entropy change  are depicted  in Fig.~\ref{fig8x}. It is quite clear that the MCE detected at $E/J\!=\!0.0$ can be significantly modulated in a proximity of  electric-field-induced phase transitions $\vert{\rm F}_+\rangle$-$\vert{\rm QF}_+\rangle$  [Fig.~\ref{fig8x}$(a)$],  $\vert{\rm QF}_-\rangle$-$\vert{\rm QF}_+\rangle$  [Fig.~\ref{fig8x}$(b)$] or both of them [Fig.~\ref{fig8x}$(c)$]. This fact can be convincingly evidenced   through the positive  isothermal entropy changes ($-\Delta S_T^E\!<\!0$) achieved upon  variation of the  electric field energy, which corroborate a possible existence of the inverse ECE. 
At each investigated case the global minimum, and thus the maximal inverse ECE, is detected exactly at the electric-field transition with the highest thermal stability, if a single electric-field driven phase transition takes place (Fig.~\ref{fig8x}$(a)$, $(b)$). In contradiction to this, the maximal temperature span of the inverse ECE in specific case with two consecutive critical points (Fig.~\ref{fig8x}$(c)$) lies in between both transition points. The most pronounced conventional ECE is evidenced at higher values of the electric field energy, but the Land\'e $g$-factor as well as the strength of the single-ion anisotropy significantly reduce the temperature of the ECE.
\\
\begin{figure*}[h!]
\centering
{\includegraphics[width=0.32\textwidth,trim=0.5cm 7.5cm 2cm 8cm, clip]{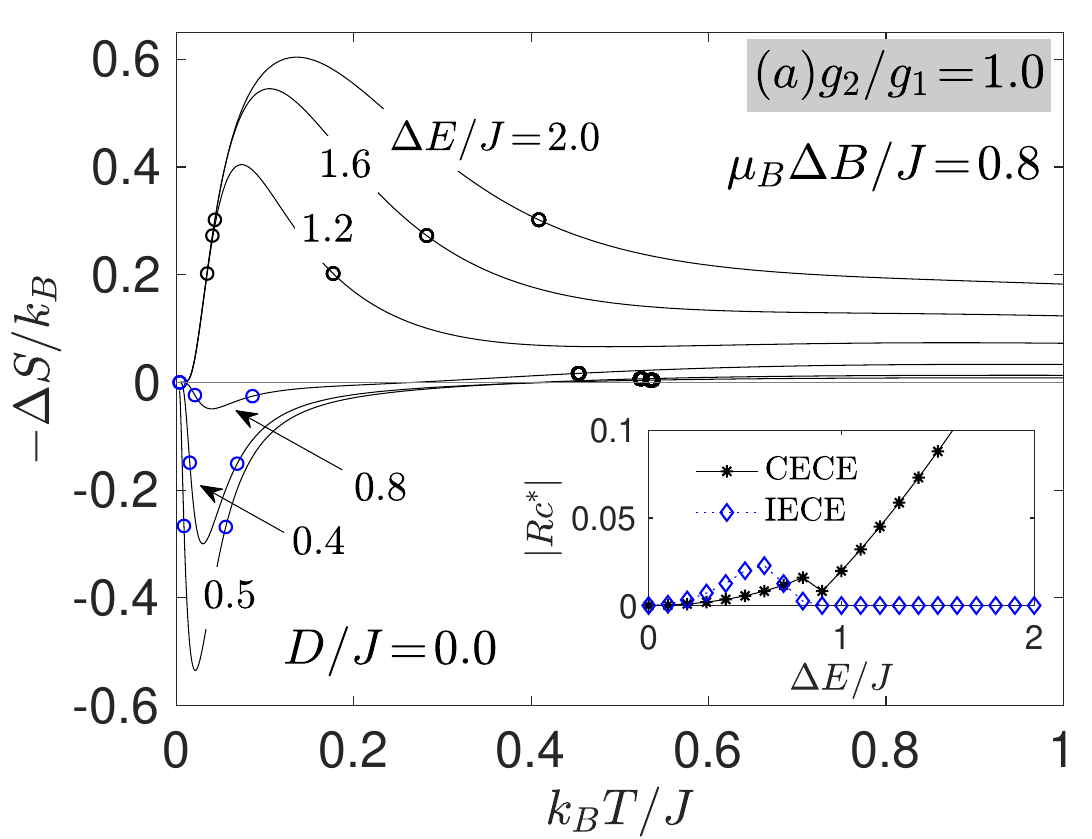}}
{\includegraphics[width=0.32\textwidth,trim=0.5cm 7.5cm 2cm 8cm, clip]{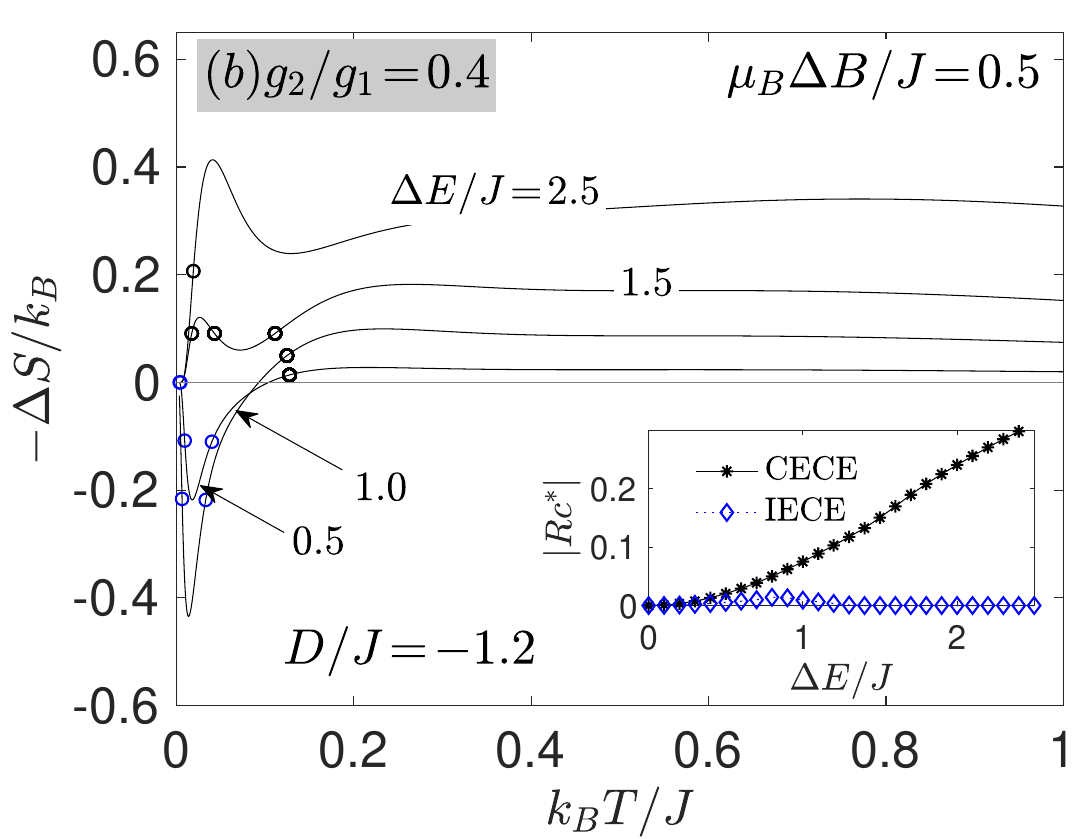}}
{\includegraphics[width=0.32\textwidth,trim=0.5cm 7.5cm 2cm 8cm, clip]{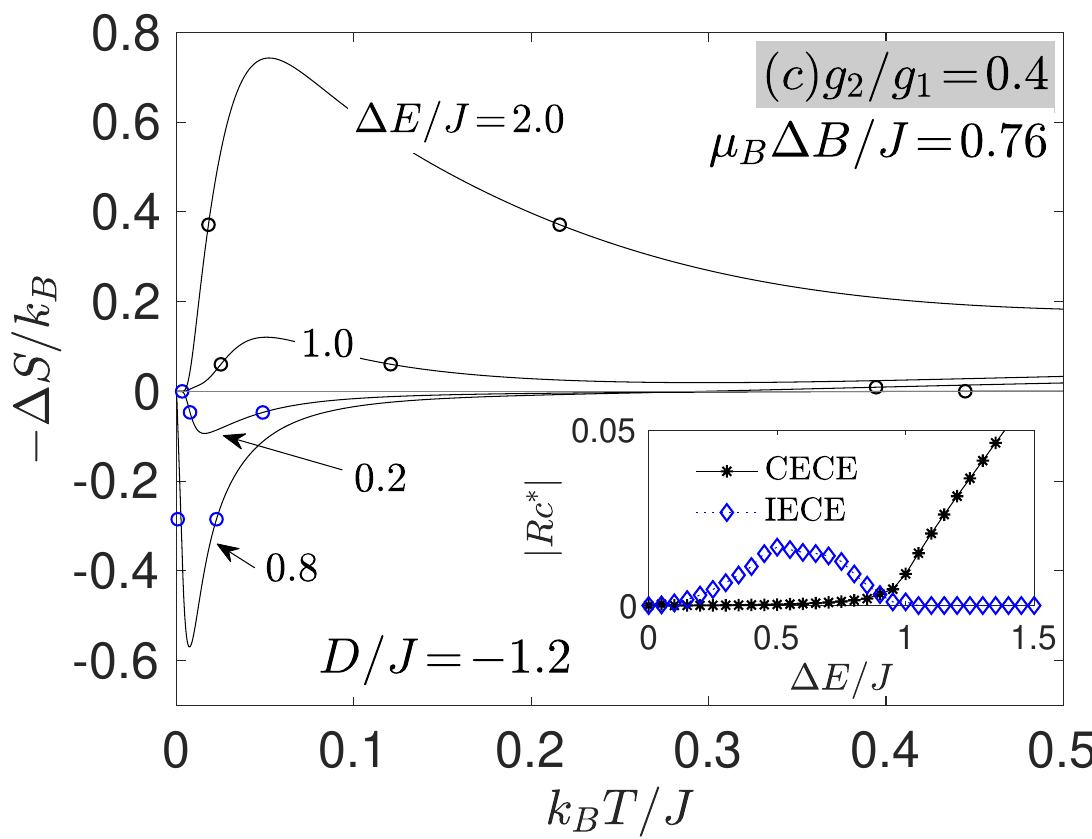}}
\caption{The dependence of a negative isothermal entropy change on the temperature for selected values of $\mu_B\Delta B/J$ and selected set of $D/J$, $g_2/g_1$ parameters. Open black and blue circles  in main panels denote limiting temperatures emerging in the Eq.~\eqref{eq25xa} to calculate the $|Rc|$ coefficient for the  conventional ECE (CECE) and inverse ECE (IECE), respectively. Insets: The evolution of the $|Rc|$ coefficient under the external magnetic field span. Black (blue) curves illustrate the change of $Rc$ between $T_1$ and $T_2$ temperature for the conventional (inverse) ECE.}
\label{fig8a}
\end{figure*}\\
The cross-section of Fig.~\ref{fig8x} clearly demonstrates that the increasing electric field energy in the region of $\vert {\rm QF}_+\rangle$ ground state significantly enhances the ECE with a motion of cooling temperature span to the higher value, see Fig.~\ref{fig8a}. In this case the magnitude of the isothermal entropy change enlarges due to the enhancement of the cooling temperature span ($-\Delta S\!>\!0$). Equally, the increment of the $Rc$ coefficient demonstrates the enhancement of the conventional ECE. Contrary to this, a sufficiently small value of electric field energy can enhance the inverse ECE, however the further increase of electric field energy completely reduces this nontrivial behavior. As a result, one can identify  a very small maximum at $|Rc|$ behavior, see insets of Fig.~\ref{fig8a}. It is interesting to note, that in the parametric space with  two consecutive electric-field transition points  the $|Rc|$ coefficient of the inverse ECE involves two more or less visible maxima, because the  highest heat transfer between reservoirs is realized in a close proximity of electric-field driven phase transition. 
\section{Conclusion}
\label{conclusion}
In the present paper we have rigorously examined the magnetocaloric and electrocaloric behavior of a mixed spin-(1/2, 1) Heisenberg dimer achieved upon  variation of the external magnetic and electric fields. An exhaustive analysis of ground state  points on  existence of the field-driven phase transitions between the quantum ferrimagnetic phases $\vert{\rm QF}_{\pm}\rangle$ and the classical ferromagnetic phase $\vert{\rm F}_{+}\rangle$ depending basically on  the uniaxial single-ion anisotropy $D/J$ and the ratio $g_2/g_1$  between Land\'e $g$-factors of both magnetic ions. It turns out that  the difference of Land\'e $g$-factors  generates an extraordinary  behavior of the total magnetization and dielectric polarization  with a pronounced quasi-plateau behavior. A mutual interaction between the magnetization and dielectric polarization serves in evidence of a  magnetoelectric effect. 
A  degeneracy accompanying   zero-temperature phase transition driven either by the external magnetic or electric field  causes   an enhancement of the caloric effect. Three basic magnetocaloric characteristics, namely the isentropic changes of temperature,  the isothermal entropy changes and the refrigerant capacity have been used to study the MCE and ECE in  a wide range of the model parameters. It was found that the mixed spin-(1/2, 1) Heisenberg dimer may exhibit a conventional as well as inverse MCE in a proximity of   magnetic-field-induced phase transitions.  It was shown, furthermore,  that  existence of an inverse MCE is  strictly conditioned by presence of the  magnetic-field-driven phase  transition between one of two  quantum ferrimagnetic phases $\vert{\rm QF}_{\pm}\rangle$ and the fully polarized ferromagnetic phase $\vert {\rm F}_+\rangle$  under the assumption of the imbalanced ratio $g_2/g_1$ between Land\'e $g$-factors of constituent spins. The most interesting observation derived from our analysis is a possibility of enhancement of the MCE through a relative low-energy-consuming way, namely,  an application of  the external electric field. Utilizing the behavior of the refrigerant capacity it was demonstrated that the increasing electric field  can significantly reduce the inverse ECE and can stabilize its conventional counterpart. As was shown furthermore,  the conventional and inverse ECE in the mixed spin-(1/2, 1) Heisenberg dimer can be additionally tuned  through the modulation of the magnetic field.  Bearing all this in mind,  the  mixed spin-(1/2, 1)   Heisenberg dimer represents a  good theoretical tool for a rigorous study of  unconventional multicaloric properties with a huge application potential. 
 \\\\
\textbf{Acknowledgments}\\
This work was financially supported by the grant of the Slovak Research and Development Agency provided under the contract No. APVV-20-0150 and by the grant of The Ministry of Education, Science, Research, and Sport of the Slovak Republic and Slovak Academy of Sciences provided under the contract No. VEGA 1/0105/20. 
\\\\
\textbf{Data Availability Statement}\\
The data that support the findings of this study are available from the corresponding author upon reasonable request.
\\\\
\textbf{Conflict of interest}\\
The authors declare that they have no known conflict of interest.
\appendix
\section*{Appendix}
\label{App A}

The exact analytical expression of the partition function of the mixed spin-(1/2,1) Heisenberg dimer
\begin{align}
{\cal Z}\!=\!\sum_{i=1}^6 {\rm e}^{-\beta \varepsilon_i}
\!=&2\left\{
\exp\left[-\frac{\beta}{2}(J+2D)\right]\cosh\left[\frac{\beta}{2}(h_1\!+\!2h_2)\right]\right.
\!+\!\exp\left[{\frac{\beta}{4}(J\!-\!2D)}\right]\left[\sum_{k=0,1}\exp\left[ (-1)^{k}\frac{\beta h_2}{2}\right]\right.
\nonumber\\
&\left.\hspace{-0.45cm}\times\!\cosh\!\left(\frac{\beta }{4}\sqrt{\left[J\!-\!2D\!-\!(-1)^{k}2(h_1\!-\!h_2)\right]^2\!\!+\!8\left[(J\Delta)^2\!+\!E^2\right]}\right)
\!\Bigg]\right\},
\label{eq11xa}
\end{align} 
where $\beta\!=\!1/(k_BT)$, $k_B$ denotes the Boltzmann's constant and $T$ is the absolute temperature. 

The exact analytical expression of normalized magnetization derived from Eq.~\eqref{eq12x}
\begin{align}
\frac{m}{m_s}&\!=\!-\frac{\partial F}{\partial \mu_BB}\!=\!\frac{1}{\cal Z}\left\{
\exp\left[-\frac{\beta}{2}(J\!+\!2D)\right](g_1\!+\!2g_2)\sinh
\left[\frac{\beta}{2}(h_1\!+\!2h_2)\right]\!+\!
\exp\left[\frac{\beta}{4}(J\!-\!2D)\right]
\right.
\nonumber\\
&\times\left[\exp\left(\frac{\beta h_{2}}{2}\right)
\left(
g_2\cosh\left[\frac{\beta }{4}\sqrt{\left[J\!-\!2D\!-\!2(h_1\!-\!h_2)\right]^2\!+\!8\left[(J\Delta)^2\!+\!E^2\right]}\right]\right.\right.
\nonumber\\
&\!-\!\left.\frac{(g_1\!-\!g_2)(J\!-\!2D\!-\!2(h_1\!-\!h_2))}{\sqrt{\left[J\!-\!2D\!-\!2(h_1\!-\!h_2)\right]^2\!+\!8\left[(J\Delta)^2\!+\!E^2\right]}}
\sinh\left[\frac{\beta }{4}
\sqrt{\left[J\!-\!2D\!-\!2(h_1\!-\!h_2)\right]^2\!+\!8\left[(J\Delta)^2\!+\!E^2\right]}
\right]\right)
\nonumber\\
&\!-\!\exp\left(-\frac{\beta h_{2}}{2}\right)\left(
g_2\cosh\left[\frac{\beta }{4}\sqrt{\left[J\!-\!2D\!+\!2(h_1\!-\!h_2)\right]^2\!+\!8\left[(J\Delta)^2\!+\!E^2\right]}\right]\right.
\nonumber\\
&\!-\!\left.\left.\left.\frac{(g_1\!-\!g_2)(J\!-\!2D\!+\!2(h_1\!-\!h_2))}{\sqrt{\left[J\!-\!2D\!+\!2(h_1\!-\!h_2)\right]^2\!+\!8\left[(J\Delta)^2\!+\!E^2\right]}}
\sinh\left[\frac{\beta }{4}
\sqrt{\left[J\!-\!2D\!+\!2(h_1\!-\!h_2)\right]^2\!+\!8\left[(J\Delta)^2\!+\!E^2\right]}
\right]\right)\right]\right\}.
\label{eq12xa}
\end{align}
The exact analytical expression of the dimensionless dielectric polarization derived from Eq.~\eqref{eq13x}
\begin{align}
P\!&=\!-{\mu}\frac{\partial F}{\partial E}\!=\!\frac{4E}{\cal Z}\exp\left[\frac{\beta}{4}(J\!-\!2D)\right]\left\{
\exp\left(\frac{\beta h_2}{2}\right)\frac{\sinh\left[\frac{\beta}{4}\sqrt{\left[J\!-\!2D\!-\!2(h_1\!-\!h_2)\right]^2\!+\!8\left[(J\Delta)^2\!+\!E^2\right]}\right]}{\sqrt{\left[J\!-\!2D\!-\!2(h_1\!-\!h_2)\right]^2\!+\!8\left[(J\Delta)^2\!+\!E^2\right]}}
\right.
\nonumber\\
&\!+\!\left. \exp\left(-\frac{\beta h_2}{2}\right)\frac{\sinh\left[\frac{\beta}{4}\sqrt{\left[J\!-\!2D\!+\!2(h_1\!-\!h_2)\right]^2\!+\!8\left[(J\Delta)^2\!+\!E^2\right]}\right]}{\sqrt{\left[J\!-\!2D\!+\!2(h_1\!-\!h_2)\right]^2\!+\!8\left[(J\Delta)^2\!+\!E^2\right]}}\right\}.
\label{eq13xa}
\end{align}

Finally, the explicit analytical expression of the entropy satisfies the Eq.~\eqref{eq14x}
\begin{align}
\displaystyle
S&=-\frac{\partial F}{\partial T}\!=\!k_B\ln {\cal Z}\!+\!
\frac{1}{T{\cal Z}}\left\{
\exp\left[-\frac{\beta}{2}(J\!+\!2D)\right]
\left[\left(J\!+\!2D\right)\cosh\left[\frac{\beta}{2}(h_1\!+\!2h_2)\right]\!-\!(h_1\!+\!2h_2)\sinh\left[\frac{\beta }{2}(h_1\!+\!2h_2)\right]\right]\right.
\nonumber\\
&
-\frac{1}{2}\exp\left[\frac{\beta}{4}(J\!-\!2D)\right]\left[\exp\left(\frac{\beta h_{2}}{2}\right)\Bigg((J\!-\!2D\!+\!2h_2)\cosh\left(\frac{\beta }{4}\sqrt{\left[J\!-\!2D\!-\!2(h_1\!-\!h_2)\right]^2\!+\!8\left[(J\Delta)^2\!+\!E^2\right]}\right)\right.
\nonumber\\
&+\sqrt{\left[J\!-\!2D\!-\!2(h_1\!-\!h_2)\right]^2\!+\!8\left[(J\Delta)^2\!+\!E^2\right]}\sinh\left(\frac{\beta}{4}\sqrt{\left[J\!-\!2D\!-\!2(h_1\!-\!h_2)\right]^2\!+\!8\left[(J\Delta)^2\!+\!E^2\right]}\right)\Bigg)
\nonumber\\
&+\exp\left(-\frac{\beta h_{2}}{2}\right)\Bigg((J\!-\!2D\!-\!2h_2)\cosh\left(\frac{\beta }{4}\sqrt{\left[J\!-\!2D
\!+\!2(h_1\!-\!h_2)\right]^2\!+\!8\left[(J\Delta)^2\!+\!E^2\right]}\right)
\nonumber\\
&\left.\left.+\sqrt{\left[J\!-\!2D\!+\!2(h_1\!-\!h_2)\right]^2\!+\!8\left[(J\Delta)^2\!+\!E^2\right]}\sinh\left(\frac{\beta}{4}\sqrt{\left[J\!-\!2D\!+\!2(h_1\!-\!h_2)\right]^2\!+\!8\left[(J\Delta)^2\!+\!E^2\right]}\right)\Bigg)
\right]\right\}
\;.
\label{eq14xa}
\end{align}
%
%

\end{document}